\documentclass[12pt]{article}

\AtBeginDocument{
  \addtocontents{toc}{\normalsize}
}
\pdfoutput=1

\usepackage{amsmath,amssymb,amsfonts,amsthm,amscd,mathrsfs}
\usepackage{xcolor}
\definecolor{darkblue}{rgb}{0.1,0.1,.7}
\usepackage[colorlinks, linkcolor=darkblue, citecolor=darkblue, urlcolor=darkblue, linktocpage]{hyperref}  
\usepackage[]{version}
\usepackage[]{graphicx}
\usepackage[]{latexsym}
\usepackage{geometry}
\usepackage[all,cmtip]{xy}
\usepackage[margin=10pt,font=small,labelfont=bf]{caption}
\usepackage{ifthen}
\usepackage{tikz}
\usepackage{array,setspace,mathrsfs,amsfonts,yfonts,dsfont,bbm,colonequals}
\usepackage{dsfont}
\usepackage{cite}
\usepackage{xspace}

\numberwithin{equation}{section}

\newcommand{\cO}{\mathcal O}

\newcommand{\be}{\begin{equation}}
\newcommand{\ee}{\end{equation}}
\newcommand{\bea}{\begin{eqnarray}}
\newcommand{\eea}{\end{eqnarray}}
\newcommand{\ba}{\begin{equation}\begin{aligned}}
\newcommand{\ea}{\end{aligned}\end{equation}}

\newcommand{\Df}{{\Delta_\phi}}

\newcommand{\Dch}{{\Delta_\chi}}

\newcommand{\email}[1]{\vbox{\center\tt#1}\vspace{5mm}}

\newcommand{\dDisc}{\mathrm{dDisc}}

\newcommand{\Gt}{\widetilde{\mathcal{G}}}
\newcommand{\sgn}{\mathrm{sgn}}

\renewcommand{\d}{\mathrm{d}}

\begin{document}

\vspace*{-.6in} \thispagestyle{empty}
\vspace{1cm} {\Large
\begin{center}
{\bf A Crossing-Symmetric OPE Inversion Formula}\\
\end{center}}
\vspace{1cm}
\begin{center}
{\bf Dalimil Maz\'a\v{c}$^{a,b}$,}\\[1cm] 
{
\small
$^{a}$ {\em C. N. Yang Institute for Theoretical Physics, Stony Brook University\\Stony Brook, NY 11794, USA}\\
$^{b}$ {\em Simons Center for Geometry and Physics, Stony Brook University\\ Stony Brook, NY 11794, USA}
\normalsize
}
\\
\end{center}

\email{dalimil.mazac@stonybrook.edu}

\vspace{4mm}

\begin{abstract}
We derive a Lorentzian OPE inversion formula for the principal series of $sl(2,\mathbb{R})$. Unlike the standard Lorentzian inversion formula in higher dimensions, the formula described here only applies to fully crossing-symmetric four-point functions and makes crossing symmetry manifest. In particular, inverting a single conformal block in the crossed channel returns the coefficient function of the crossing-symmetric sum of Witten exchange diagrams in AdS, including the direct-channel exchange. The inversion kernel exhibits poles at the double-trace scaling dimensions, whose contributions must cancel out in a generic solution to crossing. In this way the inversion formula leads to a derivation of the Polyakov bootstrap for $sl(2,\mathbb{R})$. The residues of the inversion kernel at the double-trace dimensions give rise to analytic bootstrap functionals discussed in recent literature, thus providing an alternative explanation for their existence. We also use the formula to give a general proof that the coefficient function of the principal series is meromorphic in the entire complex plane with poles only at the expected locations.
\end{abstract}
\vspace{2in}


\newpage

{
\setlength{\parskip}{0.05in}
\tableofcontents
\renewcommand{\baselinestretch}{1.0}\normalsize
}


\setlength{\parskip}{0.1in}
\newpage

\section{Introduction}\label{sec:introduction}
The conformal bootstrap is the idea of solving for the dynamics of conformal field theories starting from the basic principles of conformal invariance, unitarity and associativity of the operator product expansion. This approach has received widespread attention ever since the numerical studies of \cite{Rattazzi:2008pe,ElShowk:2012ht} demonstrated its unexpected constraining power in more than two dimensions.\footnote{See \cite{Ferrara:1973yt,Polyakov:1974gs,Mack:1975jr} for early formulations of the conformal bootstrap and \cite{Poland:2018epd} for a review of recent developments.} In parallel, an analytic approach to the conformal bootstrap has been developed, initially based on the expansion of the bootstrap equations near a pair of light-cones \cite{Komargodski:2012ek,Fitzpatrick:2012yx,Alday:2015ewa,Alday:2016njk}.

More recently, the results of the analytic conformal bootstrap have been unified and extended through the so-called Lorentzian inversion formula \cite{Caron-Huot2017b}.\footnote{See \cite{Simmons-Duffin:2017nub,Kravchuk:2018htv} for more details on and generalizations of the original Lorentzian inversion formula.} The formula exploits complex analyticity of the four-point function to extract its OPE decomposition from the double commutator, also called double discontinuity and denoted $\dDisc\!\left[\mathcal{G}(z,\bar{z})\right]$.\footnote{For identical external scalar primaries, $\mathcal{G}(z,\bar{z})$ is defined by $\langle\phi(x_1)\phi(x_2)\phi(x_3)\phi(x_4)\rangle = \langle\phi(x_1)\phi(x_2)\rangle\langle\phi(x_3)\phi(x_4)\rangle\,\mathcal{G}(z,\bar{z})$, where $x_i$ are the coordinate vectors. The cross-ratios are defined by $z\bar{z} = \frac{x_{12}^2x_{34}^2}{x_{13}^2x_{24}^2}$, $(1-z)(1-\bar{z}) = \frac{x_{14}^2x_{23}^2}{x_{13}^2x_{24}^2}$ with $x_{ij}=x_i-x_j$. The notation and conventions of this note follow closely those of \cite{Simmons-Duffin:2017nub}.} More precisely, the formula computes the coefficient function $I_{\Delta,J}$ of the decomposition of the four-point function into a complete set of conformal partial waves labelled by their dimension $\Delta$ and spin $J$. The OPE decomposition of $\mathcal{G}(z,\bar{z})$ can be read off from the poles and residues of $I_{\Delta,J}$. The Lorentzian inversion formula expresses $I_{\Delta,J}$ as an integral of the double discontinuity times an appropriate inversion kernel over a Lorentzian spacetime diamond $0<z,\bar{z}<1$.

One of the most important virtues of the Lorentzian inversion formula is that it allows one to compute the OPE data exchanged in a given channel in terms of the OPE data in the crossed channels. This in turn leads to systematic expansions of the OPE data at large spin, and large scaling dimension \cite{Mukhametzhanov:2018zja}. Furthermore, the contributions of crossed-channel operators at mean-field double-trace scaling dimensions are suppressed by the formula, allowing for a recursive determination of the OPE data in perturbation theory around the generalized free field \cite{Alday:2017zzv,Aharony:2018npf,Aharony:2016dwx,Alday:2017xua,Alday:2017vkk,Alday:2018pdi}.

The inversion formula of \cite{Caron-Huot2017b} is valid as long as the spacetime dimension is greater than one. One may be interested in having also an analogous formula which assumes only the minimal conformal symmetry, namely the global conformal symmetry of a line, corresponding to the algebra $so(1,2)=sl(2,\mathbb{R})$. One reason to look for such formula is that there exist intrinsically one-dimensional conformal-invariant systems, such as line defects in higher-D CFTs \cite{Billo:2013jda,Gaiotto:2013nva,Giombi:2018qox,Giombi:2017cqn,Giombi:2018hsx,Liendo:2018ukf} or SYK-like models \cite{Sachdev:1992fk,Maldacena:2016hyu}, to which the standard Lorentzian inversion formula in $D>1$ does not apply. Furthermore, one should keep in mind that every higher-D CFT is in particular also a 1D CFT since its correlators can be restricted to a line and satisfy all the axioms of the $sl(2,\mathbb{R})$ conformal bootstrap. Finally, 1D CFTs provide a simpler but still constraining setting for testing ideas about higher-D conformal bootstrap.

One needs to face some obvious challenges when trying to derive a Lorentzian inversion formula for $sl(2,\mathbb{R})$. First, it may not be immediately clear what the distinction is between the meaning of ``Euclidean'' and ``Lorentzian'' in one dimension. Second, the existence of the Lorentzian formula in $D>1$ is closely tied to the fact that the CFT data are analytic in spin. This property may seem mysterious from the $D=1$ point of view for the simple reason that there is no spin in one dimension. In fact, these two points have already been addressed in \cite{Simmons-Duffin:2017nub} and \cite{Maldacena:2016hyu}. To explain their resolution of the above puzzles and how the present work fits in the existing literature, let us first quickly review the basic kinematics of 1D CFTs.

We consider the four-point function of identical $sl(2,\mathbb{R})$ primaries in a unitary, parity-invariant 1D CFT. It can be written as
\be
\langle\phi(x_1)\phi(x_2)\phi(x_3)\phi(x_4)\rangle = \langle\phi(x_1)\phi(x_2)\rangle\langle\phi(x_3)\phi(x_4)\rangle\,\mathcal{G}(z)\,,
\ee
where $x_i$ are positions on the line and $z=\frac{x_{12}\,x_{34}}{x_{13}\,x_{24}}$ is the only cross-ratio of four points. A priori, the cross-ratio ranges over all real numbers but the analytic continuation to complex $z$ plays an important role in various contexts, including the present note. If the 1D four-point function arises by restricting a higher-D four-point function $\mathcal{G}_{D>1}(z,\bar{z})$ to collinear configurations, we find $\mathcal{G}(z) = \mathcal{G}_{D>1}(z,z)$. The four-point function can be expanded in a complete set of conformal partial waves using the Euclidean inversion formula. In $D>1$, this complete set consists only of the \emph{principal series} $\Delta=\frac{D}{2}+i s$ with $s\in\mathbb{R}_{+}$ and $J\in\mathbb{Z}_{\geq0}$. The complete set in $D=1$ includes both the principal series $\Delta=\frac{1}{2}+i\mathbb{R}_{+}$ and the \emph{discrete series} $\Delta\in2\mathbb{N}$.\footnote{Here and in the rest of this note, $\mathbb{N}$ stands for \emph{positive} integers and $\mathbb{R}_+$ for positive real numbers.} Correspondingly, the four-point function is described by a pair of coefficient functions: $I_{\Delta}$ for the principal series and $\widetilde{I}_{\Delta}$ for the discrete series. $I_{\Delta}$ is analogous to $I_{\Delta,J}$ from $D>1$ in the sense that primary operators in the $\phi\times\phi$ OPE with generic scaling dimensions translate into poles of $I_{\Delta}$ on the positive real axis. As we review in the main text, the main role of the discrete series is to cancel spurious contributions of the principal series.\footnote{In special circumstances, the discrete series may capture the OPE data of physical operators, see \cite{Caron-Huot:2018kta}.} The Euclidean inversion formula expresses the coefficient functions as integrals of $\mathcal{G}(z)$ over the real line as follows
\be
I_\Delta= \!\!\! \int\limits_{-\infty}^{\infty}\!\!\!dz z^{-2}\,\Psi_{\Delta}(z)\,\mathcal{G}(z)\quad\textrm{for }\Delta\in\frac{1}{2}+i\mathbb{R}\,,\quad
\widetilde{I}_\Delta= \!\!\! \int\limits_{-\infty}^{\infty}\!\!\!dz z^{-2}\,\Psi_{\Delta}(z)\,\mathcal{G}(z)\quad\textrm{for }\Delta\in2\mathbb{N}\,,
\ee
where $\Psi_{\Delta}(z)$ are the conformal partial waves. A priori, these formulas only compute the coefficient functions for $\Delta$ restricted to the principal and discrete series respectively and thus do not provide their analytic continuation to the $\Delta$ complex plane. Consequently, $I_{\Delta}=\widetilde{I}_{\Delta}$ does not hold in general.

We are interested in a Lorentzian, rather than Euclidean, inversion formula for $I_{\Delta}$ and $\widetilde{I}_{\Delta}$. Following \cite{Simmons-Duffin:2017nub}, we will take this to mean a formula which extracts the coefficient functions from the double discontinuity of $\mathcal{G}(z)$, rather than from the value of the Euclidean correlator. Such definition ensures that the 1D formula replicates the usefulness of its higher-D cousin for the analytic bootstrap. In 1D, the double discontinuity is defined by
\be
\dDisc\!\left[\mathcal{G}(z)\right] = \mathcal{G}(z)-\frac{\mathcal{G}^{\curvearrowleft}(z)+\mathcal{G}^{\text{\rotatebox[origin=c]{180}{\reflectbox{$\curvearrowleft$}}}}(z)}{2}
\quad\textrm{for }z\in(0,1)\,,
\ee
where $\mathcal{G}^{\curvearrowleft}(z)$ and $\mathcal{G}^{\text{\rotatebox[origin=c]{180}{\reflectbox{$\curvearrowleft$}}}}(z)$ are the analytic continuations of $\mathcal{G}(z)$ from $z\in(1,\infty)$ to $z\in(0,1)$ above and below the branch cut. When $\mathcal{G}(z)$ arises from a $D>1$ correlator by restricting to collinear configuration, then the above definition agrees with the standard $D>1$ double discontinuity restricted to $z=\bar{z}$. Furthermore, this definition also agrees with the thermal expectation value of the double commutator $\frac{1}{2}[\phi(x_3),\phi(x_2)][\phi(x_1),\phi(x_4)]$, where $\phi(x_3)$ and $\phi(x_4)$ are evolved in Lorentzian time. The terms $\mathcal{G}^{\curvearrowleft}(z)$ and $\mathcal{G}^{\text{\rotatebox[origin=c]{180}{\reflectbox{$\curvearrowleft$}}}}(z)$ correspond to the out-of-time-order contributions in the double commutator. The $z\rightarrow0$ limit of these terms can be used to diagnose chaos and is analogous to the Regge limit of $D>1$ \cite{Roberts:2014ifa,Maldacena:2015waa,Perlmutter:2016pkf,Maldacena:2016hyu}.

The authors of \cite{Simmons-Duffin:2017nub} derived the following Lorentzian inversion formula for the discrete series of $sl(2,\mathbb{R})$
\be
\widetilde{I}_{\Delta} = \frac{4\Gamma(\Delta)^2}{\Gamma(2\Delta)}\int\limits_{0}^{1}\!\! dz z^{-2}G_{\Delta}(z)\,\dDisc\!\left[\mathcal{G}(z)\right]\,,
\label{eq:invDiscrete0}
\ee
where $G_{\Delta}(z)$ is the 1D conformal block. This formula provides an analytic continuation of $\widetilde{I}_{\Delta}$ away from positive even integers and thereby answers the second point raised above: the analogue of analyticity in spin for $sl(2,\mathbb{R})$ is analyticity in the label of the discrete series. This is indeed needed to make the correlator bounded in the Regge limit \cite{Maldacena:2016hyu}.

The main result of this note is an analogous formula for the coefficient function of the principal series. Such formula is clearly needed for many interesting applications since $I_{\Delta}$, and not $\widetilde{I}_{\Delta}$, carries information about the spectrum in a generic OPE. The formula takes the form
\be
I_{\Delta} = 2\!\!\int\limits_{0}^{1}\!\! dz z^{-2}H_{\Delta}(z)\,\dDisc\!\left[\mathcal{G}(z)\right]\,,
\label{eq:lorInv0}
\ee
where $H_{\Delta}(z)$ is an appropriate inversion kernel. We will fix $H_{\Delta}(z)$ by demanding compatibility of \eqref{eq:lorInv0} with the Euclidean inversion formula. For the formula to be valid, $\mathcal{G}(z)$ must be bounded in the Regge limit.

While \eqref{eq:lorInv0} is similar to the standard higher-D inversion formula, there are some key differences between the two. Most notably, the present formula only works for four-point functions of identical external operators. At the practical level, this is because the contour-deformation argument relating \eqref{eq:lorInv0} to the Euclidean inversion formula is only valid provided $\mathcal{G}(z)$ is fully Bose- or Fermi-symmetric.\footnote{On the other hand, Caron-Huot's formula works for four-point functions of arbitrary sets of external operators and indeed Bose symmetry plays no role in its derivation.} Correspondingly, we will have one formula for identical bosons and one for identical fermions. The two cases are almost identical, but the bosonic one has some additional subtleties, which we suppress in the introduction. In the fermionic case, the rest of the introduction applies without any amendment.

At first sight, the requirement of Bose/Fermi symmetry may seem like a limitation, but it implies that \eqref{eq:lorInv0} leads to an interesting reformulation of the crossing equation, as we explain in the next few paragraphs.

Suppose we start from a four-point function $\mathcal{G}(z)$ of identical primaries in a unitary theory and apply \eqref{eq:lorInv0} to it. Just like in $D>1$, $\dDisc\!\left[\mathcal{G}(z)\right]$ can be computed by expanding $\mathcal{G}(z)$ in the t-channel. The first step is then to understand the coefficient function obtained by applying the Lorentzian inversion formula to a single t-channel conformal block of general dimension $\Delta_{\cO}$, which we denote as follows
\be
\mathcal{I}(\Delta;\Delta_{\mathcal{O}}|\Df) \equiv
2\!\!\int\limits_{0}^{1}\!\! dz z^{-2}H_{\Delta}(z)\,\dDisc\!\left[\left(\mbox{$\frac{z}{1-z}$}\right)^{2\Df}G_{\Delta_{\mathcal{O}}}(1-z)\right]\,.
\label{eq:iCal0}
\ee
We will argue in the main text that $\mathcal{I}(\Delta;\Delta_{\mathcal{O}}|\Df)$ describes the crossing-symmetric sum of exchange Witten diagrams in $AdS_2$ in the s-, t- and u-channel with exchanged dimension $\Delta_{\mathcal{O}}$. We call this crossing-symmetric object the \emph{Polyakov block} for reasons that will become clear soon. The Polyakov block with exchanged dimension $\Delta_{\mathcal{O}}$ will be denoted $P_{\Delta_{\mathcal{O}}}(z)$. The s-channel OPE decomposition of $P_{\Delta_{\mathcal{O}}}(z)$ contains the single-trace conformal block of dimension $\Delta_{\mathcal{O}}$ with unit coefficient, as well as an infinite tower of double-trace contributions with dimensions $2\Df+2n$ or $2\Df+2n+1$ in the bosonic and fermionic case respectively, where $n=0,1,\ldots$. In summary, using \eqref{eq:lorInv0} to invert a single block in the crossed channel returns a manifestly crossing-symmetric object. This is in contrast with what happens when using the standard $D>1$ Lorentzian inversion formula, as described in Section \ref{sec:discussion} of this note.

Having understood the contribution of an individual crossed-channel block to $I_{\Delta}$, we will argue that one can commute the integral in \eqref{eq:lorInv0} and the t-channel OPE applied to $\dDisc\!\left[\mathcal{G}(z)\right]$. This means $I_{\Delta}$ can be expanded using the coefficient functions of the Polyakov blocks as follows
\be
I_{\Delta} = \!\!\sum\limits_{\mathcal{O}\in\phi\times\phi}\!(c_{\phi\phi\mathcal{O}})^2\,\mathcal{I}(\Delta;\Delta_{\mathcal{O}}|\Df)\,,
\ee
where the sum runs over $sl(2,\mathbb{R})$ primaries in the $\phi\times\phi$ OPE and $c_{\phi\phi\mathcal{O}}$ are the OPE coefficients. The sum converges (absolutely and uniformly in any compact set) in the entire complex $\Delta$-plane away from poles of the individual terms in the sum. Since $\mathcal{I}(\Delta;\Delta_{\mathcal{O}}|\Df)$ are meromorphic functions of $\Delta$, it follows that $I_{\Delta}$ is also meromorphic, with poles only at locations of poles of the individual terms. The same expansion then holds also at the level of the correlator
\be
\mathcal{G}(z) = \!\!\sum\limits_{\mathcal{O}\in\phi\times\phi}\!(c_{\phi\phi\mathcal{O}})^2 P_{\Delta_{\mathcal{O}}}(z)\,.
\label{eq:polExp0}
\ee
On the other hand, we know $\mathcal{G}(z)$ can be expanded in the s-channel conformal blocks as follows
\be
\mathcal{G}(z) = \!\!\sum\limits_{\mathcal{O}\in\phi\times\phi}\!(c_{\phi\phi\mathcal{O}})^2 G_{\Delta_{\mathcal{O}}}(z)\,.
\label{eq:gExp0}
\ee
For the last two equations to be compatible, the double-trace contributions to the Polyakov blocks must cancel out when the sum over $\mathcal{O}$ in \eqref{eq:polExp0} has been performed. This leads to an infinite set of sum rules on the OPE data, with two independent sum rules for every double-trace operator. This is precisely the idea behind Polyakov's approach to the conformal bootstrap \cite{Polyakov:1974gs} recently revisited and refined using Mellin-space techniques in \cite{Sen:2015doa,Gopakumar2017,Gopakumar2017a}.

These sum rules were recently derived and studied in a closely related work \cite{Mazac:2018ycv}. Among other things, reference \cite{Mazac:2018ycv} demonstrated not only that \eqref{eq:polExp0} holds for every unitary solution of $sl(2,\mathbb{R})$ crossing, but also the stronger claim that the totality of sum rules arising from the equivalence of \eqref{eq:polExp0} and \eqref{eq:gExp0} is in fact completely equivalent to the standard crossing equation
\be
z^{-2\Df}\!\!\sum\limits_{\mathcal{O}\in\phi\times\phi}\!(c_{\phi\phi\mathcal{O}})^2 G_{\Delta_{\mathcal{O}}}(z) = 
(1-z)^{-2\Df}\!\!\sum\limits_{\mathcal{O}\in\phi\times\phi}\!(c_{\phi\phi\mathcal{O}})^2 G_{\Delta_{\mathcal{O}}}(1-z)\,.
\label{eq:crossing0}
\ee
One goal of the present note is to offer an alternative perspective on the same core idea from the point of view of the Lorentzian inversion formula.

In \cite{Mazac:2018ycv}, the relevant sum rules were derived by applying suitable linear functionals to the crossing equation \eqref{eq:crossing0}, building on the constructions of \cite{Mazac:2016qev,Mazac:2018mdx}. The functionals themselves are interesting because they are examples of optimal (or extremal) functionals of the numerical bootstrap \cite{ElShowk:2012hu}. For example, they can be used to show rigorously that the gap above identity in the $\phi\times\phi$ OPE in a unitary CFT (in any $D$) is at most $2\Df+1$, where $\phi$ is a scalar primary or a component of a spinning primary. The bound becomes optimal if only the minimal (1D) conformal symmetry is assumed. The most important properties of the functionals is that when acting on the conformal block (minus the crossed-channel conformal block) of dimension $\Delta_{\mathcal{O}}$, they are positive from a certain $\Delta_{\mathcal{O}}$ onwards and exhibit double zeros for $\Delta_{\mathcal{O}}$ at the double-trace values. The functionals of \cite{Mazac:2016qev,Mazac:2018mdx,Mazac:2018ycv} are constructed as contour integrals against suitable holomorphic kernels in the complex $z$-plane. The kernels are constrained by an intricate functional equation which guarantees the above properties.

In this note, we will derive this construction starting from the Lorentzian inversion formula \eqref{eq:lorInv0}. The Lorentzian inversion kernel $H_{\Delta}(z)$ has double poles for $\Delta$ at the double-trace dimensions. These poles precisely reproduce the double-trace contributions to the Polyakov block in formula \eqref{eq:iCal0}. It turns out that the coefficients of the simple and double poles of $H_{\Delta}(z)$ at the double traces are precisely the kernels used to define the functionals of \cite{Mazac:2016qev,Mazac:2018mdx,Mazac:2018ycv}. The contour integral prescription for the functionals of these works is nothing but a way to reconstruct the double discontinuity while staying on the first sheet in the $z$ variable. The intricate functional equation satisfied by the functional kernels is a consequence of the equation satisfied by $H_{\Delta}(z)$ which guarantees that the Lorentzian and Euclidean inversion formulas give the same answer for $I_{\Delta}$. The functionals exhibit positivity and double zeros at the double traces because the double discontinuity of conformal blocks in the crossed channel has these properties. In this way, the inversion formula of this note unifies all the functionals considered in \cite{Mazac:2016qev,Mazac:2018mdx,Mazac:2018ycv} into a single object.

\subsection*{Outline and summary of results}
The rest of this note is structured as follows. In Section \ref{sec:kinematics}, we review 1D kinematics and the expansion of a four-point function into a complete set of conformal partial waves provided by the Euclidean inversion formula.

In Section \ref{sec:LorentzianInversion}, we discuss the Lorentzian inversion formula for the principal series and explain how the inversion kernel is constrained by compatibility with the Euclidean formula.

We find explicit formulas for the Lorentzian inversion kernel in Section \ref{sec:HFormulas}. This includes closed formulas for the bosonic kernel when $\Df\in\mathbb{N}$ and the fermionic kernel when $\Df\in\mathbb{N}-\frac{1}{2}$. Furthermore, we find the series expansion of the kernel around $z=0$ for general $\Df$.

Section \ref{sec:witten} explains that the Lorentzian inverse of a single conformal block in the t-channel is the coefficient function of a fully crossing-symmetric sum of exchange Witten diagrams in $AdS_2$, including the s-channel exchange.

The implications of the last observation are analyzed in Section \ref{sec:polyakov}. We prove that $I_{\Delta}$ of a crossing-symmetric four-point function in a unitary theory can be expanded in the coefficient functions of crossing-symmetrized exchange Witten diagrams. We explain why this implies $I_{\Delta}$ is meromorphic in the entire complex plane with poles only at the expected locations. Furthermore, we explain how consistency with the usual OPE leads to infinitely many sum rules on the CFT data. Finally, we demonstrate that optimal functionals of the numerical bootstrap arise from residues of the Lorentzian inversion kernel at the double-trace locations.

We conclude with a discussion and open questions in Section \ref{sec:discussion}.

\section{Kinematics and the Euclidean inversion formula}\label{sec:kinematics}
\subsection{The four-point function}
In this note, we will focus on the four-point function of identical operators $\phi(x)$ in a conformal field theory, denoted $\langle\phi(x_1)\phi(x_2)\phi(x_3)\phi(x_4)\rangle$. Let us restrict the four operators to lie on a straight line in the Euclidean space and let $x$ denote the coordinate along the line. There is a conformal symmetry $sl(2,\mathbb{R})$ acting along this line. We will take $\phi(x)$ to be a primary operator of dimension $\Df$ with respect to this $sl(2,\mathbb{R})$. Thus $\phi(x)$ can be for example a scalar primary operator or a component of a spinning operator in a $D>1$ CFT, or simply a primary operator of a 1D CFT. In the rest of the note, we will distinguish the two cases where $\phi(x)$ has bosonic and fermionic statistics. Let us focus on the bosonic case first. The two-point function then reads
\be
\langle\phi(x_1)\phi(x_2)\rangle = \frac{1}{|x_{12}|^{2\Df}}\,,
\ee
where $x_{ij} = x_i-x_j$. Symmetry under $sl(2,\mathbb{R})$ implies that the four-point function can be written as
\be
\langle\phi(x_1)\phi(x_2)\phi(x_3)\phi(x_4)\rangle = \langle\phi(x_1)\phi(x_2)\rangle\langle\phi(x_3)\phi(x_4)\rangle\,\mathcal{G}(z)\,,
\ee
where $z$ is the cross-ratio
\be
z = \frac{x_{12}x_{34}}{x_{13}x_{24}}\,.
\ee
We can use the conformal symmetry and a permutation of labels 1 and 3 if necessary to set $x_1=0$, $x_3=1$ and $x_4=\infty$. $z$ is then equal to $x_2$ and thus ranges over all real numbers. When $\langle\phi(x_1)\phi(x_2)\phi(x_3)\phi(x_4)\rangle$ arises by restricting a $D>1$ four-point function to a line, $\mathcal{G}(z)$ is obtained by restricting the full four-point function $\mathcal{G}(z,\bar{z})$ to $\bar{z}=z$.

Since the four-point function $\mathcal{G}(z)$ is singular at coincident points $z=0,1,\infty$, it is useful to define $\mathcal{G}^{(-)}(z)$, $\mathcal{G}^{(0)}(z)$ and $\mathcal{G}^{(+)}(z)$ as the functions to which $\mathcal{G}(z)$ reduces in the three disconnected regions
\ba
\mathcal{G}(z) =
\begin{cases}
\mathcal{G}^{(-)}(z)\quad&\textrm{for }z\in(-\infty,0)\\
\mathcal{G}^{(0)}(z)\quad&\textrm{for }z\in(0,1)\\
\mathcal{G}^{(+)}(z)\quad&\textrm{for }z\in(1,\infty)\,.
\end{cases}
\label{eq:GCases}
\ea
The functions $\mathcal{G}^{(0,\pm)}(z)$ can be analytically continued to complex values of $z$, but in general are not analytic continuations of each other. Instead, they can be related by Bose symmetry of the four-point function. The symmetry under $1\leftrightarrow 2$ determines $\mathcal{G}^{(-)}(z)$ in terms of $\mathcal{G}^{(0)}(z)$ and the symmetry under $2\leftrightarrow3$ determines $\mathcal{G}^{(+)}(z)$ in terms of $\mathcal{G}^{(0)}(z)$ as follows
\ba
\mathcal{G}^{(-)}(z) &= \mathcal{G}^{(0)}\!\left(\mbox{$\frac{z}{z-1}$}\right)\quad\;\;\;\,\textrm{for } z\in(-\infty,0)\\
\mathcal{G}^{(+)}(z) &= z^{2\Df}\mathcal{G}^{(0)}\!\left(\mbox{$\frac{1}{z}$}\right)\quad\textrm{for } z\in(1,\infty)\,.
\label{eq:gCalRelations}
\ea
Clearly $\mathcal{G}^{(0)}(z)$ determines the whole four-point function. In addition, symmetry under $2\leftrightarrow 4$ implies $\mathcal{G}^{(0)}(z)$ must satisfy the crossing relation
\be
z^{-2\Df}\mathcal{G}^{(0)}(z) = (1-z)^{-2\Df}\mathcal{G}^{(0)}(1-z)\quad\textrm{for } z\in(0,1)\,.
\ee
It will be convenient to define the function
\be
\widetilde{\mathcal{G}}(z) = z^{-2\Df}\mathcal{G}^{(0)}(z)\,,
\ee
for which crossing symmetry becomes $\widetilde{\mathcal{G}}(z) = \widetilde{\mathcal{G}}(1-z)$.

The four-point function $\mathcal{G}^{(0)}(z)$ can be expanded using the s-channel OPE $\phi(x_1)\times\phi(x_2)$. Since we are only assuming $sl(2,\mathbb{R})$ symmetry, the appropriate conformal blocks are the $sl(2,\mathbb{R})$ blocks
\be
G_{\Delta}(z) = z^{\Delta}{}_2F_1(\Delta,\Delta;2\Delta;z)\,.
\label{eq:1Dblock}
\ee
It will also be useful to define the conformal block for negative $z$
\be
\widehat{G}_{\Delta}(z)\equiv G_{\Delta}\!\left(\mbox{$\frac{z}{z-1}$}\right) =  (-z)^{\Delta}{}_2F_1(\Delta,\Delta;2\Delta;z)\,.
\ee
The s-channel OPE reads
\be
\mathcal{G}^{(0)}(z) = \!\!\sum\limits_{\mathcal{O}\in\phi\times\phi}\!\!(c_{\phi\phi\mathcal{O}})^2 G_{\Delta_{\mathcal{O}}}(z)\,,
\ee
where the sum runs over the primary operators appearing in the OPE and $c_{\phi\phi\mathcal{O}}$ is the appropriate OPE coefficient. A priori, the expansion holds for $z\in(0,1)$ but a standard argument using the $\rho$ coordinate \cite{Hogervorst:2013sma,Rychkov:2017tpc} shows that the sum converges also for complex $z$ away from $z\in[1,\infty)$. The conformal blocks $G_{\Delta}(z)$ have a power-law branch cut at $(-\infty,0]$ implying $\mathcal{G}^{(0)}(z)$ has branch cuts for $z\in(-\infty,0]$ and $z\in[1,\infty)$. Let us also note that thanks to an asymptotic bound on OPE coefficients \cite{Pappadopulo:2012jk,Qiao:2017xif}, the convergence of the s-channel OPE is uniform in any compact region of $\mathbb{C}$ not containing $[1,\infty)$, implying $\mathcal{G}^{(0)}(z)$ is holomorphic away from the branch cuts $(-\infty,0]$ and $[1,\infty)$.

We will also need the knowledge of how $\mathcal{G}^{(0)}(z)$ behaves for large $z$. In order to avoid the branch cuts, we should take the limit in the upper half-plane\footnote{The lower half-plane being related by crossing $z\leftrightarrow 1-z$.} as $z=r e^{i\theta}$, $r\in\mathbb{R}$ and $r\rightarrow\infty$. This limit of the four-point function is precisely the Regge limit of the u-channel, as explained in detail in Section 2 of \cite{Mazac:2018ycv}. It is also the same limit that can be used to diagnose chaos from out-of-time-order correlators \cite{Roberts:2014ifa,Maldacena:2015waa,Perlmutter:2016pkf}. Since all channels are equivalent for identical external operators, we will simply refer to this limit as the Regge limit. We will also take $z\rightarrow\infty$ to mean approaching $\infty$ in any direction in the upper (or lower) half-plane. Four-point functions in unitary theories satisfy a boundedness condition in the Regge limit. To see that, one can work in the $\rho$ coordinate and use positivity of the s-channel expansion together with the fact that the t-channel is dominated by the identity operator \cite{Rychkov:2017tpc}. The result is that
\be
|\widetilde{\mathcal{G}}(z)|\textrm{ is bounded as }z\rightarrow\infty\,.
\ee
Note that the bound can not be improved in general since $\widetilde{\mathcal{G}}(z)$ of the generalized free field approaches a constant in the Regge limit
\be
\widetilde{\mathcal{G}}(z) = z^{-2\Df}+(1-z)^{-2\Df}+1\rightarrow 1
\ee
as $z\rightarrow \infty$. For technical reasons, it will sometimes be useful to consider functions which are better-behaved than just bounded in the Regge limit. Thus, let us define $\widetilde{\mathcal{G}}(z)$ to be \emph{super-bounded} if
\be
|\widetilde{\mathcal{G}}(z)| = O(|z|^{-1-\epsilon})\textrm{ as }z\rightarrow\infty
\label{eq:superbounded}
\ee
for $\epsilon>0$.

Consider now the case of identical fermions $\chi(x)$. The two-point function has an extra ordering sign
\be
\langle\chi(x_1)\chi(x_2)\rangle = \frac{\sgn(x_{12})}{|x_{12}|^{2\Dch}}\,.
\ee
The four-point function $\mathcal{G}(z)$ is defined analogously to the bosonic case
\be
\langle\chi(x_1)\chi(x_2)\chi(x_3)\chi(x_4)\rangle = \langle\chi(x_1)\chi(x_2)\rangle\langle\chi(x_3)\chi(x_4)\rangle\,\mathcal{G}(z)\,.
\label{eq:GFermDef}
\ee
The functions $\mathcal{G}^{(0,\pm)}(z)$ are defined exactly as in \eqref{eq:GCases}. Symmetry under the permutations of the external operators again fixes $\mathcal{G}^{(-)}(z)$ and $\mathcal{G}^{(+)}(z)$ in terms of $\mathcal{G}^{(0)}(z)$. The transposition $1\leftrightarrow2$ introduces an ordering sign on both sides of \eqref{eq:GFermDef}, which thus cancel and give
\be
\mathcal{G}^{(-)}(z) = \mathcal{G}^{(0)}\!\left(\mbox{$\frac{z}{z-1}$}\right)\quad\;\;\;\,\textrm{for } z\in(-\infty,0)\,.
\ee
On the other hand, the transposition $2\leftrightarrow3$ leads to an extra sign compared to the bosonic case
\be
\mathcal{G}^{(+)}(z) = - z^{2\Dch}\mathcal{G}^{(0)}\!\left(\mbox{$\frac{1}{z}$}\right)\quad\textrm{for } z\in(1,\infty)\,.
\ee
We will again define $\widetilde{\mathcal{G}}(z) = z^{-2\Dch}\mathcal{G}^{(0)}(z)$. Just like in the bosonic case, $\widetilde{\mathcal{G}}(z)$ satisfies crossing symmetry $\widetilde{\mathcal{G}}(z) = \widetilde{\mathcal{G}}(1-z)$ and boundedness in the Regge limit (the latter whenever the theory is unitary).

\subsection{Review of the Euclidean inversion formula}
The Plancherel theorem for $SL(2,\mathbb{R})$ allows us to expand $\mathcal{G}(z)$ into a complete set of eigenfunctions of the s-channel Casimir. Both in the bosonic and fermionic case, $\mathcal{G}(z)$ is invariant under $z\leftrightarrow\mbox{$\frac{z}{z-1}$}$ for $z\in\mathbb{R}$. Since the s-channel Casimir respects the same symmetry, we can restrict to eigenfunctions invariant under this symmetry. Moreover, the eigenfunctions need to satisfy a boundary condition at $z=1$ to ensure the Casimir operator is self-adjoint, see for example Section 3.2.2 of \cite{Maldacena:2016hyu} for more details. The relevant eigenfunctions are called conformal partial waves and can be written as the following conformal integral\footnote{The integral representation converges for $0<\mathrm{Re}(\Delta)<1$. For other values of $\Delta$, $\Psi_{\Delta}(z)$ can be defined by an analytic continuation from this region.}
\be
\Psi_{\Delta}(z) = \int\limits_{-\infty}^{\infty}\!\!dx_5\left(\frac{|x_{12}|}{|x_{15}||x_{25}|}\right)^{\Delta}
\left(\frac{|x_{34}|}{|x_{35}||x_{45}|}\right)^{1-\Delta}\quad\textrm{for }z\in\mathbb{R}\,.
\label{eq:psiIntRep}
\ee
For $z\in(0,1)$, the conformal partial waves are a linear combination of a conformal block and its shadow
\be
\Psi^{(0)}_{\Delta}(z) = K_{1-\Delta}G_{\Delta}(z) + K_{\Delta}G_{1-\Delta}(z)\,,
\label{eq:psi0}
\ee
where
\be
K_{\Delta} =
\frac{\sqrt{\pi }\Gamma \left(\Delta -\frac{1}{2}\right) \Gamma \left(\frac{1-\Delta }{2}\right)^2}{\Gamma (1-\Delta ) \Gamma \left(\frac{\Delta }{2}\right)^2}\,.
\ee
For $z\in(-\infty,0)$ and $z\in(1,\infty)$, the conformal partial waves are determined from $\Psi^{(0)}_{\Delta}(z)$ as follows
\ba
\Psi^{(-)}_{\Delta}(z) &= \Psi^{(0)}_{\Delta}\left(\mbox{$\frac{z}{z-1}$}\right)\qquad\qquad\qquad\qquad\qquad\;\;\,
\textrm{for }z\in(-\infty,0)\\
\Psi^{(+)}_{\Delta}(z) &=\frac{1}{2}\left[\Psi^{(0)}_{\Delta}(z+i\epsilon)+\Psi^{(0)}_{\Delta}(z-i\epsilon)\right]
\qquad\textrm{for }z\in(1,\infty)\,.
\label{eq:psiRelations}
\ea
The invariant inner product on $sl(2,\mathbb{R})$ gives the following inner product of functions of $z$
\be
\left(\mathcal{G}_1,\mathcal{G}_2\right) =\!\!\! \int\limits_{-\infty}^{\infty}\!\!\!dz z^{-2}\,\mathcal{G}_1(z)\,\mathcal{G}_2(z)\,.
\label{eq:innerproduct}
\ee
The set of conformal partial waves which is orthogonal and complete with respect to this inner product consists of the principal series $\Delta = 1/2+ i\alpha$ with $\alpha\in\mathbb{R}_{+}$ and the discrete series $\Delta\in 2\mathbb{N}$. Note that on the discrete series, the second term in \eqref{eq:psi0} vanishes and we find
\be
\Psi^{(0)}_{m}(z) = \frac{2\Gamma(m)^2}{\Gamma(2m)}G_{m}(z)\quad\textrm{for }m\in2\mathbb{N}\,.
\ee

The scalar products among the complete set are
\ba
&\left( \Psi_{\frac{1}{2}+i\alpha},\Psi_{\frac{1}{2}+i\beta}\right) = 2\pi n_{\frac{1}{2}+i\alpha}\delta(\alpha-\beta)\quad\quad\, \alpha,\beta\in\mathbb{R}_+\\
&\left( \Psi_{m},\Psi_{n}\right) =\frac{4\pi^2}{2m-1}\delta_{mn}\qquad\qquad\qquad\quad\;\; m,n\in 2\mathbb{N}\\
&\left( \Psi_{\frac{1}{2}+i\alpha},\Psi_{m}\right) = 0\qquad\qquad\qquad\qquad\qquad\; \alpha\in\mathbb{R}_+,\,m\in2\mathbb{N}\,,
\ea
where
\be
n_{\Delta} = 2 K_{\Delta}K_{1-\Delta} = \frac{4\pi\tan(\pi\Delta)}{2\Delta-1}\,.
\ee
Let us define the following coefficient functions as overlaps of the four-point function with the principal and discrete series conformal partial waves
\ba
I_\Delta &= \left(\Psi_{\Delta},\mathcal{G}\right)=\!\!\! \int\limits_{-\infty}^{\infty}\!\!\!dz z^{-2}\,\Psi_{\Delta}(z)\,\mathcal{G}(z)
\quad\textrm{for }\Delta=\frac{1}{2}+i\alpha,\,\alpha\in\mathbb{R}\\
\widetilde{I}_m &= \left(\Psi_{m},\mathcal{G}\right)=\!\!\! \int\limits_{-\infty}^{\infty}\!\!\!dz z^{-2}\,\Psi_{m}(z)\,\mathcal{G}(z)
\quad\textrm{for }m\in2\mathbb{N}\,.
\label{eq:euclInv}
\ea
The four-point function can then be expanded in the complete set as follows
\be
\mathcal{G}(z) =\!\!\int\limits_{\frac{1}{2}}^{\frac{1}{2} + i\infty}\!\frac{d\Delta}{2\pi i}\frac{I_{\Delta}}{n_{\Delta}}\Psi_{\Delta}(z) + \sum\limits_{m\in2\mathbb{N}}\frac{2m-1}{4\pi^2}\widetilde{I}_{m}\Psi_{m}(z)\,.
\ee
For $z\in(0,1)$, we can use \eqref{eq:psi0} and $I_{\Delta} = I_{1-\Delta}$ to write this as
\be
\mathcal{G}(z) =\!\!\int\limits_{\frac{1}{2} - i\infty}^{\frac{1}{2} + i\infty}\!\frac{d\Delta}{2\pi i}\frac{I_{\Delta}}{2K_{\Delta}}G_{\Delta}(z) + \sum\limits_{m\in2\mathbb{N}}\frac{\Gamma (m)^2}{2 \pi ^2 \Gamma (2 m-1)}\widetilde{I}_m G_{m}(z)\,.
\label{eq:opeFromIs}
\ee
If $\mathcal{G}(z)$ is normalizable with respect to the inner product \eqref{eq:innerproduct}, $I_{\Delta}$ is holomorphic in some neighbourhood of the principal series, and $\widetilde{I}_m$ is finite for $m\in 2\mathbb{N}$. Note that it may not always be true that $\widetilde{I}_m$ is the analytic continuation of $I_{\Delta}$ from the principal series to $\Delta=m$ since the integral defining $I_{\Delta}$ may not converge along the path of the analytic continuation.

The s-channel OPE is recovered by closing the contour to the right, so that terms of the OPE come from poles of $I_{\Delta}/K_{\Delta}$ and the terms of the discrete series sum. Concretely, we can define
\be
c(\Delta) = \frac{I_{\Delta}}{2K_{\Delta}}
\ee
so that the presence of $\mathcal{O}$ in the $\phi\times\phi$ OPE translates to a simple pole of $c(\Delta)$ at $\Delta = \Delta_{\mathcal{O}}$ with residue $-(c_{\phi\phi\mathcal{O}})^2$. There is no reason for the $\phi\times\phi$ OPE to always contain operators with scaling dimensions exactly at even integers, so how should we think of the contribution of the discrete series? The answer comes from noting that $K_{\Delta}$ has simple zeros on the discrete series, leading to potentially unphysical poles of $c(\Delta)$. When there is no physical operator at $\Delta=m\in2\mathbb{N}$, we have $I_{m}=\widetilde{I}_m$. This guarantees that the residue of the integral in \eqref{eq:opeFromIs} coming from the zero of $K_{\Delta}$ at $\Delta=m$ precisely cancels the corresponding term of the sum over the discrete series. On the other hand, if there is a physical operator precisely at $\Delta=m\in2\mathbb{N}$, we should have $I_{m}\neq\widetilde{I}_m$ so that the principal series integral and the discrete series sum combine to a non-vanishing contribution at $\Delta=m$.

We have seen that the formulas \eqref{eq:euclInv} extract the OPE data from the Euclidean four-point function and for this reason are known as the Euclidean inversion formulas. To see some concrete examples, we can consider the four-point functions of the generalized free boson (GFB) and fermion (GFF). In the bosonic case, the four-point function is
\be
\mathcal{G}(z) = 1 + \left|\mbox{$\frac{z}{z-1}$}\right|^{2\Df} + |z|^{2\Df}\,.
\ee
The inversion integral \eqref{eq:euclInv} is most easily done by using the integral representation of the conformal partial waves \eqref{eq:psiIntRep}. The final result is\footnote{Strictly speaking, to compute $I^{\textrm{GFB}}_{\Delta}$ one needs to remove the non-normalizable contribution of the identity and work with $\mathcal{G}(z) = \left|\mbox{$\frac{z}{1-z}$}\right|^{2\Df} + |z|^{2\Df}$ instead.}
\be
I^{\textrm{GFB}}_{\Delta} = \frac{\pi \Gamma \left(\Df-\frac{\Delta }{2}\right)\Gamma \left(\Df+\frac{\Delta-1}{2}\right)\Gamma \left(\frac{\Delta+2}{2}-\Df\right)\Gamma \left(\frac{3-\Delta }{2}-\Df\right)}{\cos ^2(\pi  \Df) \Gamma (2 \Df)^2 \Gamma (\Delta +1-2\Df)\Gamma (2-\Delta-2 \Df)}\,.
\ee
$I^{\textrm{GFB}}_{\Delta}$ is essentially the simplest meromorphic function with the right poles and residues which respects the shadow symmetry. The four-point function of the generalized free fermion of dimension $\Dch$ is
\be
\mathcal{G}(z) = 1 -\sgn\left(\mbox{$\frac{z}{z-1}$}\right) \left|\mbox{$\frac{z}{z-1}$}\right|^{2\Dch} -\sgn(z) |z|^{2\Dch}\,.
\ee
Note that in spite of the fermionic statistics, we have $\mathcal{G}(z) = +\mathcal{G}\left(\mbox{$\frac{z}{z-1}$}\right)$ and we can use the same set of conformal partial waves as in the bosonic case. The coefficient function reads
\be
I^{\textrm{GFF}}_{\Delta} =
-\frac{\pi  \Gamma\left(\Dch-\frac{\Delta-1}{2}\right)\Gamma \left(\Dch+\frac{\Delta }{2}\right)\Gamma \left(\frac{\Delta+1}{2}-\Dch\right)\Gamma \left(\frac{2-\Delta}{2}-\Dch\right)}{\sin ^2(\pi  \Dch) \Gamma (2\Dch)^2\Gamma (\Delta +1-2\Dch) \Gamma (2-\Delta-2\Dch)}\,.
\ee

Before closing this section, let us note that if we are using the Euclidean inversion formula to extract the OPE data of a four-point function restricted to $0<z<1$, we can be agnostic about the statistics of the external operators. Indeed, suppose we are given $\mathcal{G}(z)$ for $0<z<1$ satisfying $z^{-2\Df}\mathcal{G}(z) = (1-z)^{-2\Df}\mathcal{G}(1-z)$ and we want to find coefficient functions $I_{\Delta}$, $\widetilde{I}_{m}$ such that \eqref{eq:opeFromIs} holds. In order to compute the inversion integrals \eqref{eq:euclInv}, we need to extend $\mathcal{G}(z)$ to a function defined for all $z\in\mathbb{R}$, such that $\mathcal{G}(z) = \mathcal{G}\left(\mbox{$\frac{z}{z-1}$}\right)$. One way to do this is to pretend the external operators are identical bosons, which gives
\ba
\mathcal{G}^{\textrm{B}}(z) =
\begin{cases}
\mathcal{G}\left(\mbox{$\frac{z}{z-1}$}\right)\quad&\textrm{for }z\in(-\infty,0)\\
\mathcal{G}(z)\quad&\textrm{for }z\in(0,1)\\
z^{2\Df}\mathcal{G}\left(\mbox{$\frac{1}{z}$}\right)\quad&\textrm{for }z\in(1,\infty)\,.
\end{cases}
\label{eq:extensionB}
\ea
Another option is to pretend they are identical fermions\footnote{From now on, we will call the external dimension $\Df$ also in the fermionic case to simplify notation.}
\ba
\mathcal{G}^{\textrm{F}}(z) =
\begin{cases}
\mathcal{G}\left(\mbox{$\frac{z}{z-1}$}\right)\quad&\textrm{for }z\in(-\infty,0)\\
\mathcal{G}(z)\quad&\textrm{for }z\in(0,1)\\
-z^{2\Df}\mathcal{G}\left(\mbox{$\frac{1}{z}$}\right)\quad&\textrm{for }z\in(1,\infty)\,.
\end{cases}
\label{eq:extensionF}
\ea
Both are perfectly consistent choices which lead to \emph{different} coefficient functions $I^{\textrm{B}}_{\Delta}$ and $I^{\textrm{F}}_{\Delta}$. $I^{\textrm{B}}_{\Delta}$ and $I^{\textrm{F}}_{\Delta}$ encode the same OPE data pertaining to the original correlator $\mathcal{G}(z)$ and thus must have the same residues at the physical poles. Therefore, their difference must be a meromorphic function with poles only at $\Delta=1,3,\ldots$ (and their shadow locations) since these poles (but not their shadows) cancel against zeros of $1/K_{\Delta}$ in \eqref{eq:opeFromIs} and thus do not contribute to the OPE. The main result of this note are alternative formulas which extract $I^{\textrm{B}}_{\Delta}$ and $I^{\textrm{F}}_{\Delta}$ from $\mathcal{G}(z)$, to which we turn now.

\section{The Lorentzian inversion formula}\label{sec:LorentzianInversion}
\subsection{The general form}
For many applications, it is essential to have an alternative formula for the coefficient functions $I_{\Delta}$ and $\widetilde{I}_m$, known as the Lorentzian inversion formula \cite{Caron-Huot2017b} (see also \cite{Simmons-Duffin:2017nub,Kravchuk:2018htv}). The input of this formula is the double discontinuity of $\mathcal{G}(z)$ defined by
\be
\dDisc\!\left[\mathcal{G}(z)\right] = \mathcal{G}^{(0)}(z)-\frac{\mathcal{G}^{(+)}(z+i\epsilon)+\mathcal{G}^{(+)}(z-i\epsilon)}{2}
\quad\textrm{for }z\in(0,1)\,.
\ee
When $\mathcal{G}(z)$ is obtained by restricting a higher-D four-point function to $z=\bar{z}$, then this definition agrees with the standard higher-D double discontinuity restricted to $z=\bar{z}$. 

All s-channel conformal partial waves are annihilated by the double discontinuity. Crucially for many applications, the double discontinuity also annihilates t-channel double-trace conformal blocks and their derivatives with respect to $\Delta$. Which dimensions are counted as double-trace depends on whether we choose the bosonic or fermionic extension of the $0<z<1$ correlator, i.e. equations \eqref{eq:extensionB} or \eqref{eq:extensionF}. In the bosonic case, the contribution of a t-channel conformal block of dimension $\Delta$ to $\mathcal{G}^{(0)}(z)$ and $\mathcal{G}^{(+)}(z)$ is
\ba
\mathcal{G}^{(0)}(z) &=\left(\mbox{$\frac{z}{1-z}$}\right)^{2\Df}G_{\Delta}(1-z)\\
\mathcal{G}^{(+)}(z) &=\left(\mbox{$\frac{z}{z-1}$}\right)^{2\Df}\widehat{G}_{\Delta}(1-z)\,,
\ea
which means its double discontinuity takes the form
\be
\dDisc_{\textrm{B}}\!\left[ G^{(t)}_{\Delta}(z)\right]=2\sin^2\!\left[\frac{\pi}{2}(\Delta-2\Df)\right]\left(\mbox{$\frac{z}{1-z}$}\right)^{2\Df}G_{\Delta}(1-z)\,,
\ee
where the subscript on $\dDisc$ reminds us which extension of the correlator we choose. We see that the bosonic double discontinuity is non-negative for $0<z<1$ and exhibits double zeros at the bosonic double-trace dimensions
\be
\Delta^{\textrm{B}}_n = 2\Df + 2n\,,\quad n=0,1,\ldots\,.
\ee
The contribution of a t-channel conformal block in the fermionic case is
\ba
\mathcal{G}^{(0)}(z) &=\left(\mbox{$\frac{z}{1-z}$}\right)^{2\Df}G_{\Delta}(1-z)\\
\mathcal{G}^{(+)}(z) &=-\left(\mbox{$\frac{z}{z-1}$}\right)^{2\Df}\widehat{G}_{\Delta}(1-z)\,,
\ea
leading to the following double discontinuity
\be
\dDisc_{\textrm{F}}\!\left[ G^{(t)}_{\Delta}(z)\right]=2\cos^2\!\left[\frac{\pi}{2}(\Delta-2\Df)\right]\left(\mbox{$\frac{z}{1-z}$}\right)^{2\Df}G_{\Delta}(1-z)\,,
\ee
this time exhibiting double zeros at the fermionic double-trace dimensions
\be
\Delta^{\textrm{F}}_n = 2\Df + 2n +1\,,\quad n=0,1,\ldots\,.
\ee

The authors of \cite{Simmons-Duffin:2017nub} found the following Lorentzian inversion formula for the discrete series coefficient function
\be
\widetilde{I}_m = \frac{4\Gamma(m)^2}{\Gamma(2m)}\int\limits_{0}^{1}\!\! dz z^{-2}G_{m}(z)\dDisc\!\left[\mathcal{G}(z)\right]\,.
\label{eq:invDiscrete}
\ee
This formula applies to any physical four-point function satisfying $\mathcal{G}(z) = \mathcal{G}\!\left(\mbox{$\frac{z}{z-1}$}\right)$, and in particular both the bosonic and fermionic extension of a crossing-symmetric $\mathcal{G}(z)$. \eqref{eq:invDiscrete} provides a particular analytic continuation of $\widetilde{I}_m$ to $m\notin2\mathbb{N}$. As discussed in \cite{Simmons-Duffin:2017nub}, this analytic continuation is holomorphic for $\mathrm{Re}(m)>1/2$ and therefore in general can not agree with the principal series function $I_{\Delta}$.

Our goal in this section will be to derive a similar formula for the principal series coefficient function $I_{\Delta}$. More precisely, we will find Lorentzian inversion formulas for $I^{\textrm{B}}_{\Delta}$, $I^{\textrm{F}}_{\Delta}$, i.e. the coefficient functions corresponding to the bosonic and fermionic extensions of $\mathcal{G}(z)$. The formulas take the form
\ba
I^{\textrm{B}}_{\Delta} &= 2\!\!\int\limits_{0}^{1}\!\! dz z^{-2}H^{\textrm{B}}_{\Delta}(z)\dDisc_{\textrm{B}}\!\left[\mathcal{G}(z)\right]\\
I^{\textrm{F}}_{\Delta} &= 2\!\!\int\limits_{0}^{1}\!\! dz z^{-2}H^{\textrm{F}}_{\Delta}(z)\dDisc_{\textrm{F}}\!\left[\mathcal{G}(z)\right]\,.
\label{eq:invPrincipal}
\ea
Here $H^{\textrm{B}}_{\Delta}(z)$ and $H^{\textrm{F}}_{\Delta}(z)$ are appropriate inversion kernels and we took out a factor of 2 for future convenience. Unlike in higher dimensions or for the discrete series, we will find that $H^{\textrm{B,F}}_{\Delta}(z)$ are \emph{not} eigenfunctions of the s-channel Casimir. In fact, they depend very non-trivially on the external dimension $\Df$. Another unusual feature of the formula is that it has the crossing symmetry under $z\leftrightarrow 1-z$ built in, in the sense that it only holds for $\mathcal{G}(z)$ respecting this symmetry. Furthermore, the output of the formula is a coefficient function $I_{\Delta}$ which manifestly leads to a crossing-symmetric correlator.\footnote{The higher-D inversion formula does not (and should not) always return crossing-symmetric OPE data. Indeed, if $\phi_{1,2}$ are scalar primaries of equal dimension, the exact same higher-D inversion formula applies to the correlators $\langle \phi_1\phi_1\phi_1\phi_1\rangle$ and $\langle\phi_1\phi_1\phi_2\phi_2\rangle$. The latter is in general not symmetric under the $s\leftrightarrow t$ crossing transformation. Our formula for the principal series only applies to fully symmetric correlators such as $\langle \phi_1\phi_1\phi_1\phi_1\rangle$.}

\subsection{Constraining the inversion kernels}\label{ssec:LorConstraint}
\begin{figure}[ht!]%
\begin{center}
\includegraphics[width=16cm]{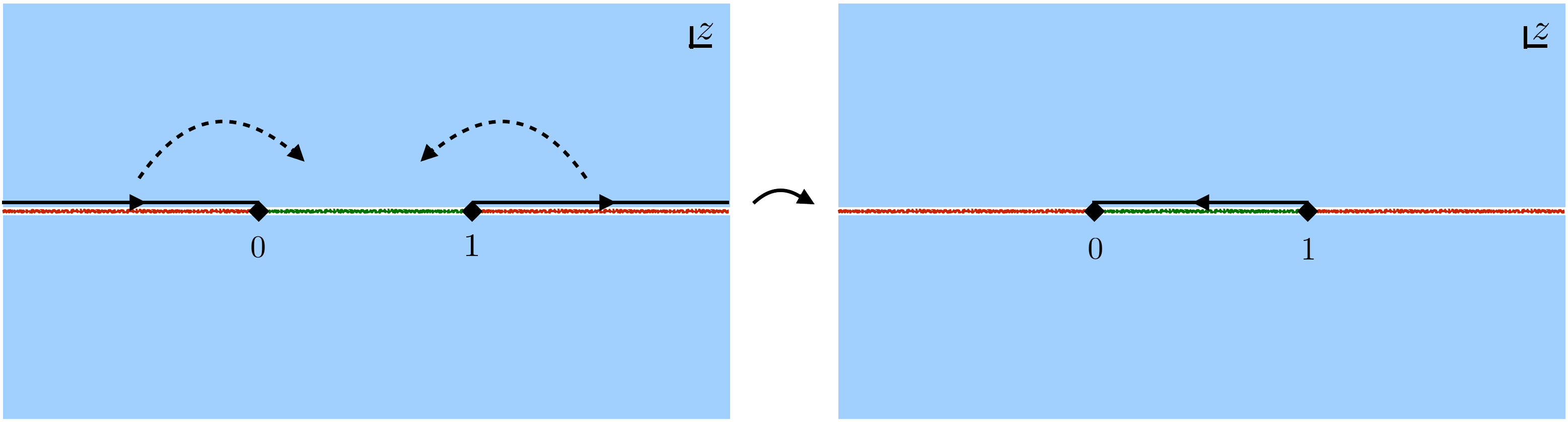}%
\caption{The contour deformation bringing the ``top-of-the-branch-cut'' contributions of the Lorentzian inversion formula to the Euclidean region. The left figure depicts the second and third term on the RHS of equation \eqref{eq:lInvInterm}. The two semi-infinite contours can be combined with a semi-circular contour at infinity and collapsed on top of the interval $z\in(0,1)$, giving the RHS of equation \eqref{eq:contourDef}, as depicted in the figure on the right. The branch cuts of $\Gt(z)$ are shown in red and the branch cut of $H_{\Delta}\!\left(\mbox{$\frac{1}{z}$}\right)$ in green.}
\label{fig:ContourDef}%
\end{center}
\end{figure}

We will fix $H^{\textrm{B}}_{\Delta}(z)$ and $H^{\textrm{F}}_{\Delta}(z)$ by demanding that the Euclidean and Lorentzian inversion formulas \eqref{eq:euclInv} and \eqref{eq:invPrincipal} are compatible. Let us start from the Euclidean formula which we first split into integrals over the three regions
\be
I^{}_{\Delta} = \!\!\!\int\limits_{-\infty}^{0}\!\!\!dz z^{-2}\Psi^{(-)}_\Delta(z)\mathcal{G}^{(-)}(z)
+\!\!\int\limits_{0}^{1}\!\!dz z^{-2}\Psi^{(0)}_\Delta(z)\mathcal{G}^{(0)}(z)
+\!\!\int\limits_{1}^{\infty}\!\!dz z^{-2}\Psi^{(+)}_\Delta(z)\mathcal{G}^{(+)}(z)\,.
\label{eq:euclInv2}
\ee
Recall that $\mathcal{G}^{(-)}(z)$ and $\mathcal{G}^{(+)}(z)$ are related to $\mathcal{G}^{(0)}(z)$ by \eqref{eq:extensionB} and \eqref{eq:extensionF} respectively in the bosonic and fermionic case. Similarly, $\Psi_{\Delta}^{(-)}(z)$ and $\Psi_{\Delta}^{(+)}(z)$ are related to $\Psi_{\Delta}^{(0)}(z)$ through \eqref{eq:psiRelations}. Let us plug these relations into \eqref{eq:euclInv2} and change variables to bring all integrations to $z\in(0,1)$. We arrive at
\be
I^{\textrm{B,F}}_{\Delta} = \!\int\limits_{0}^{1}\!\!dz\left[2z^{2\Df-2}\Psi^{(0)}_\Delta(z)\pm\frac{\Psi^{(0)}_\Delta\left(\mbox{$\frac{1}{z}$}+i\epsilon\right)+\Psi^{(0)}_\Delta\left(\mbox{$\frac{1}{z}$}-i\epsilon\right)}{2}\right]\Gt(z)\,.
\ee
Here and in the following the upper sign applies for $I^{\textrm{B}}_{\Delta}$ and the lower for $I^{\textrm{F}}_{\Delta}$. Recall that $\Gt(z) = z^{-2\Df}\mathcal{G}^{(0)}(z)$ is crossing symmetric but the square bracket in the last formula is not, so let us symmetrize it to get a formula manifesting the full crossing symmetry\footnote{The partial waves satisfy $\Psi^{(+)}(z) = \Psi^{(+)}\left(\mbox{$\frac{z}{z-1}$}\right)$, or equivalently $\Psi^{(0)}_\Delta\left(\mbox{$\frac{1}{z}$}+i\epsilon\right)+\Psi^{(0)}_\Delta\left(\mbox{$\frac{1}{z}$}-i\epsilon\right)=\Psi^{(0)}_\Delta\left(\mbox{$\frac{1}{1-z}$}+i\epsilon\right)+\Psi^{(0)}_\Delta\left(\mbox{$\frac{1}{1-z}$}-i\epsilon\right)$ for $z\in(0,1)$ so that no symmetrization on the last two terms in the square bracket is necessary.}
\ba
I^{\textrm{B,F}}_{\Delta} = \!\int\limits_{0}^{1}\!\!dz&\left[z^{2\Df-2}\Psi^{(0)}_\Delta(z)+(1-z)^{2\Df-2}\Psi^{(0)}_\Delta(1-z) \pm\phantom{\frac{\Psi^{(0)}_\Delta\left(\mbox{$\frac{1}{z}$}\right)}{2}}\right.\\
&\left.\;\;\;\;\;\;\;\;\;\;\;\;\pm\frac{\Psi^{(0)}_\Delta\left(\mbox{$\frac{1}{z}$}+i\epsilon\right)+\Psi^{(0)}_\Delta\left(\mbox{$\frac{1}{z}$}-i\epsilon\right)}{2}\right]\Gt(z)\,.
\label{eq:euclInvSimp}
\ea
The next step is to manipulate the proposed Lorentzian formulas \eqref{eq:invPrincipal} into the same form, i.e. an integral over $z\in(0,1)$ of a crossing-symmetric kernel multiplying $\Gt(z)$. Let us start by expressing the double discontinuity in terms of $\Gt(z)$
\be
\dDisc_{\textrm{B,F}}\!\left[\mathcal{G}(z)\right] = z^{2\Df}\Gt(z)\mp\frac{\Gt\left(\mbox{$\frac{1}{z}$}+i\epsilon\right)+\Gt\left(\mbox{$\frac{1}{z}$}-i\epsilon\right)}{2}\,,
\label{eq:dDiscDef}
\ee
where $z\in(0,1)$. Let us plug this in the inversion formula \eqref{eq:invPrincipal} and change coordinates so that $\Gt(z)$ only appears with argument $z$ or $z+i\epsilon$
\ba
I^{\textrm{B,F}}_{\Delta} &= \!\int\limits_{0}^{1}\!\!dzz^{-2}\,H^{\textrm{B,F}}_{\Delta}(z)\!\left[2z^{2\Df}\Gt(z) \mp\Gt\left(\mbox{$\frac{1}{z}$}-i\epsilon\right)\mp\Gt\left(\mbox{$\frac{1}{z}$}+i\epsilon\right)
\right] = \\
&=\!\int\limits_{0}^{1}\!\!dz\,2z^{2\Df-2}H^{\textrm{B,F}}_{\Delta}(z)\,\Gt(z) \mp
\!\int\limits_{-\infty}^{0}\!\!\!dz\,H^{\textrm{B,F}}_{\Delta}\left(\mbox{$\frac{1}{1-z}$}\right)\,\Gt(z+i\epsilon)\mp\\
&\;\;\;\mp\!\int\limits_{1}^{\infty}\!\!dz\,H^{\textrm{B,F}}_{\Delta}\left(\mbox{$\frac{1}{z}$}\right)\,\Gt(z+i\epsilon)\,,
\label{eq:lInvInterm}
\ea
where we used $\Gt(z-i\epsilon) = \Gt(1-z+i\epsilon)$ to manipulate the second term. For the Lorentzian formula to be compatible with the Euclidean one \eqref{eq:euclInvSimp}, we must be able to bring all the integrals to $z\in(0,1)$. The first term is already in this form. The second and third term can be brought to such form assuming $H_{\Delta}(z)$ satisfies a few requirements. We want to combine these two terms with a semi-circular contour at infinity and deform the contour to the interval $z\in(0,1)$, as shown in Figure \ref{fig:ContourDef}. Recall that $\Gt(z)$ is holomorphic in the upper half-plane. A priori, $H_{\Delta}\!\left(\mbox{$\frac{1}{1-z}$}\right)$ is defined for $z\in(-\infty,0)$ and $H_{\Delta}\!\left(\mbox{$\frac{1}{z}$}\right)$ is defined for $z\in(1,\infty)$. For the contour deformation to be allowed, the analytic continuations of these two functions to the upper half-plane must be holomorphic, and in fact equal to the same function! This observation shows that $H_{\Delta}\!\left(z\right)$ must admit a single-valued analytic continuation from $z\in(0,1)$ to $z\in\mathbb{C}\backslash[1,\infty]$, which satisfies
\be
H_{\Delta}\left(z\right) = H_{\Delta}\left(\mbox{$\frac{z}{z-1}$}\right)\quad\textrm{for }z\in\mathbb{C}\backslash[1,\infty)\,.
\label{eq:HSym}
\ee

\noindent It also follows that $H_{\Delta}(z)$ must be holomorphic away from a possible branch cut at $z\in[1,\infty)$ and a possible singularity at $z=0$. Finally, for the contour deformation of Figure \ref{fig:ContourDef} to be allowed, we must ensure the semicircle at $z=\infty$ gives a vanishing contribution. As discussed in Section \ref{sec:kinematics}, $z\rightarrow\infty$ is the Regge limit and $\Gt(z)$ goes to zero or a finite constant there. Therefore, the contribution from infinity can be dropped if $H_{\Delta}(z) = O(z^{1+\epsilon})$ as $z\rightarrow 0$ for some $\epsilon>0$. Since $H_{\Delta}(z)$ is holomorphic in an open neighbourhood of $z=0$, we now see it must in fact be holomorphic at $z=0$ too, which combined with $H_{\Delta}(z) = O(z^{1+\epsilon})$ implies $H_{\Delta}(z) = O(z^2)$ as $z\rightarrow0$. Expressed in formulas, the contour deformation gives
\ba
\!\int\limits_{-\infty}^{0}\!\!\!dz\,H_{\Delta}\left(\mbox{$\frac{1}{1-z}$}\right)\,&\Gt(z+i\epsilon)+
\!\int\limits_{1}^{\infty}\!\!dz\,H_{\Delta}\left(\mbox{$\frac{1}{z}$}\right)\,\Gt(z+i\epsilon) = 
-\!\int\limits_{0}^{1}\!\!dz\,H_{\Delta}\left(\mbox{$\frac{1}{z}$}+i\epsilon\right)\,\Gt(z)=\\
&=-\!\int\limits_{0}^{1}\!\!dz\frac{H_{\Delta}\left(\mbox{$\frac{1}{z}$}+i\epsilon\right)+H_{\Delta}\left(\mbox{$\frac{1}{z}$}-i\epsilon\right)}{2}\Gt(z)\,.
\label{eq:contourDef}
\ea
Let us proceed in the derivation by combining the contributions in \eqref{eq:lInvInterm}, using \eqref{eq:contourDef} and symmetrizing with respect to $z\leftrightarrow1-z$. This reduces the Lorentzian formulas to
\ba
I^{\textrm{B,F}}_{\Delta} &= \!\int\limits_{0}^{1}\!\!dz\left[z^{2\Df-2}H^{\textrm{B,F}}_\Delta(z)+(1-z)^{2\Df-2}H^{\textrm{B,F}}_\Delta(1-z)\pm\phantom{\frac{\Psi^{(0)}_\Delta\left(\mbox{$\frac{1}{z}$}\right)}{2}}\right.\\
&\left.\qquad\qquad\qquad\pm\frac{H^{\textrm{B,F}}_\Delta\left(\mbox{$\frac{1}{z}$}+i\epsilon\right)+H^{\textrm{B,F}}_\Delta\left(\mbox{$\frac{1}{z}$}-i\epsilon\right)}{2}\right]\Gt(z)\,.
\label{eq:invPrincipalSimp}
\ea
The result looks identical to the Euclidean formula \eqref{eq:euclInvSimp} with the simple replacement $\Psi^{(0)}_{\Delta}\mapsto H_{\Delta}$. The Euclidean and Lorentzian formulas agree for a general crossing-symmetric correlator only if the expressions in the square brackets agree for $z\in(0,1)$
\ba
&z^{2\Df-2}H^{\textrm{B,F}}_\Delta(z)+(1-z)^{2\Df-2}H^{\textrm{B,F}}_\Delta(1-z) \pm\frac{H^{\textrm{B,F}}_\Delta\left(\mbox{$\frac{1}{z}$}+i\epsilon\right)+H^{\textrm{B,F}}_\Delta\left(\mbox{$\frac{1}{z}$}-i\epsilon\right)}{2} =\\
=&z^{2\Df-2}\Psi^{(0)}_\Delta(z)+(1-z)^{2\Df-2}\Psi^{(0)}_\Delta(1-z) \pm\frac{\Psi^{(0)}_\Delta\left(\mbox{$\frac{1}{z}$}+i\epsilon\right)+\Psi^{(0)}_\Delta\left(\mbox{$\frac{1}{z}$}-i\epsilon\right)}{2}\,.
\label{eq:feIH}
\ea

A simple way to solve this equation would be to set $H_{\Delta}(z)=\Psi^{(0)}_{\Delta}(z)$. However, this function is not holomorphic in $z\in\mathbb{C}\backslash[1,\infty)$ for $\Delta$ on the principal series because of a branch point at $z=0$. Furthermore, this choice of $H_{\Delta}(z)$ does not satisfy the other necessary constraints, namely \eqref{eq:HSym} and the condition $H_{\Delta}(z)=O(z^2)$ as $z\rightarrow 0$.

It turns out that \eqref{eq:feIH} has nontrivial solutions which also satisfy all the other constraints, making the Lorentzian inversion formula possible. We were able to find the solution in many cases, building on the connection to the extremal functionals explained later. Closed formulas for the inversion kernels will be presented in the following section. The solution for general $\Delta$ and $\Df$ seems rather complicated, with a nontrivial dependence on $\Df$. This is in contrast to the higher-D inversion formula, where the inversion kernel is simply an s-channel conformal block with Weyl-reflected quantum numbers. The nontrivial $\Df$ dependence is the price we need to pay for having an inversion formula which manifests the crossing symmetry of the correlator, since $\Df$ enters the crossing equation.

Before closing this section, note that for special values of $\Delta$ the naive solution $H_{\Delta}(z)=\Psi^{(0)}_{\Delta}(z)$ \emph{does} satisfy all the additional constraints. This is precisely the discrete series $\Delta\in2\mathbb{N}$. Indeed, we will find
\be
H^{\textrm{B,F}}_{\Delta}(z) = \Psi^{(0)}_{\Delta}(z) = \frac{2\Gamma(\Delta)^2}{\Gamma(2\Delta)}G_{\Delta}(z)\quad\textrm{for }\Delta\in2\mathbb{N},\;\Df\textrm{ generic}\,.
\label{eq:phiReduction}
\ee
This reproduces the inversion formula \eqref{eq:invDiscrete} of reference \cite{Simmons-Duffin:2017nub} for the discrete series. The simplification only occurs if the double-trace dimensions $\Delta^{\textrm{B}}_n$ or $\Delta^{\textrm{F}}_n$ do not overlap with the discrete series, hence the requirement of generic $\Df$.

\section{Explicit formulas for the inversion kernels}\label{sec:HFormulas}
In this section, we will solve the constraints described above to find the Lorentzian inversion kernels $H_\Delta(z)$. Recall that the constraints are the symmetry property \eqref{eq:HSym}, the functional equation \eqref{eq:feIH} and finally the requirement $H_{\Delta}(z) = O(z^2)$ as $z\rightarrow 0$. In fact, the solution that we will present in the bosonic case will only satisfy the weaker condition $H^{\textrm{B}}_{\Delta}(z) = O(z^{0})$ as $z\rightarrow 0$. This means that at face value, the resulting bosonic inversion formula will only apply to super-bounded correlators, as defined in \eqref{eq:superbounded}. This restriction will be fixed in \ref{ssec:reggeRelax}. On the other hand, the fermionic inversion kernel constructed in the present section will satisfy $H^{\textrm{F}}_{\Delta}(z) = O(z^{2})$ as $z\rightarrow 0$, so no further amendments will be needed.

The bosonic kernel $H^{\textrm{B}}_\Delta(z)$ simplifies for $\Df\in\mathbb{N}$, whereas $H^{\textrm{F}}_\Delta(z)$ simplifies for $\Df\in\mathbb{N}-\frac{1}{2}$. These are the cases when the double-trace operators lie on the discrete series. In this section, we will describe the general form of $H^{\textrm{B,F}}_\Delta(z)$ for general $\Delta$ and these discrete values of $\Df$. Furthermore, we will present the Taylor expansions of $H^{\textrm{B,F}}_\Delta(z)$ around $z=0$ for general $\Delta$ and \emph{general} $\Df$.

\subsection{The bosonic case}
In order to describe $H^{\textrm{B}}_\Delta(z)$ for $\Df\in\mathbb{N}$, it will be useful to work with
\be
p_{\Delta}(z) = {}_2F_1(\Delta,1-\Delta;1;z)\,.
\ee
$p_{\Delta}(z)$ is holomorphic for $z\in\mathbb{C}\backslash[1,\infty)$. Furthermore, $p_{\Delta}\left(\mbox{$\frac{z-1}{z}$}\right)$ and $p_{\Delta}\left(\mbox{$\frac{1}{z}$}\right)$ are eigenfunctions of the s-channel Casimir with the same eigenvalue as the partial wave $\Psi_{\Delta}(z)$.\footnote{In fact, $p_{\Delta}\left(\mbox{$\frac{z-1}{z}$}\right)$ are precisely the basis functions of the alpha-space expansion of \cite{Hogervorst2017a}.} Therefore, we should be able to express $\Psi^{(0)}_{\Delta}(z)$ as a linear combination of $p_{\Delta}\left(\mbox{$\frac{z-1}{z}$}\right)$ and $p_{\Delta}\left(\mbox{$\frac{1}{z}$}\right)$. Indeed, the following identities hold
\ba
&\Psi^{(0)}_{\Delta}(z) = \frac{2\pi}{\sin(\pi\Delta)}\left[p_{\Delta}\left(\mbox{$\frac{z-1}{z}$}\right) + \frac{p_{\Delta}\left(\mbox{$\frac{1}{z}$}+i\epsilon\right)+p_{\Delta}\left(\mbox{$\frac{1}{z}$}-i\epsilon\right)}{2}\right]\quad\textrm{for }z\in(0,1)\\
&\frac{\Psi^{(0)}_{\Delta}(z+i\epsilon)+\Psi^{(0)}_{\Delta}(z-i\epsilon)}{2} =\frac{2\pi}{\sin(\pi\Delta)}\left[p_{\Delta}\left(\mbox{$\frac{z-1}{z}$}\right) +p_{\Delta}\left(\mbox{$\frac{1}{z}$}\right)\right]\quad\,\textrm{for }z\in(1,\infty)\,.
\ea
These identities allow us to find the inversion kernel $H^{\textrm{B}}_\Delta(z)$ for $\Df\in\mathbb{N}$. The simplest case is $\Df=1$, where we find
\be
H^{\textrm{B}}_\Delta(z)= \frac{2\pi}{\sin(\pi \Delta)}\left[p_\Delta\!\left(z\right) + p_\Delta\!\left(\mbox{$\frac{z}{z-1}$}\right)\right]\qquad(\Df=1)\,.
\label{eq:phi1}
\ee
Indeed, upon substituting this $H^{\textrm{B}}_\Delta(z)$ into \eqref{eq:feIH}, we can group all terms into three sets corresponding to eigenfunctions of the s-, t- and u-channel Casimir. The above identities ensure that the terms cancel within each set. Note that $H^{\textrm{B}}_\Delta(z) = H^{\textrm{B}}_\Delta\!\left(\mbox{$\frac{z}{z-1}$}\right)$ as required by \eqref{eq:HSym} and  $H^{\textrm{B}}_\Delta(z) = H^{\textrm{B}}_{1-\Delta}(z)$ as expected from the shadow symmetry of $I_{\Delta}$. As stated above, $H^{\textrm{B}}_\Delta(z)$ is not an eigenfunction of the s-channel Casimir. Instead, it is a symmetric combination of eigenfunctions of the t- and u-channel Casimirs. Also note that $H^{\textrm{B}}_\Delta(z)$ has poles on the double-trace dimensions $\Delta^{\textrm{B}}_n$, a fact that will be important for the connection to the Polyakov bootstrap and extremal functionals. Finally, $H^{\textrm{B}}_\Delta(z) = O(z^0)$ as $z\rightarrow0$ and thus the resulting inversion formula applies to all super-bounded $\Gt(z)$ as promised.

Moving on to higher integer values of $\Df$, we find that the combination
\be
\frac{2\pi}{\sin(\pi \Delta)}\left[z^{-2\Df+2}p_\Delta\!\left(z\right) + \left(\mbox{$\frac{z}{z-1}$}\right)^{-2\Df+2}\!\!p_\Delta\!\left(\mbox{$\frac{z}{z-1}$}\right)\right]
\label{eq:HBAns}
\ee
solves the constraint \eqref{eq:HSym} and the functional equation \eqref{eq:feIH} for all $\Df\in\mathbb{Z}$. It is nevertheless not the right answer since it has a pole at $z=0$ for $\Df>1$. We found that there is always a solution of the equation \eqref{eq:feIH} with zero RHS which can be added to \eqref{eq:HBAns} to completely cancel its poles at $z=0$.\footnote{These homogeneous solutions come in two forms. Firstly, the residues of \eqref{eq:HBAns} at $\Delta = 2,4,\ldots,2\Df-2$ clearly solve \eqref{eq:feIH} with zero RHS since the RHS has no poles there. Secondly, the difference of the finite part of \eqref{eq:HBAns} at $\Delta = 2,4,\ldots,2\Df-2$ and $ \frac{2\Gamma(\Delta)^2}{\Gamma(2\Delta)}G_{\Delta}(z)$ provides the remaining homogeneous solutions.}

In summary, we find the following general form for $\Df\in\mathbb{N}$
\be
H^{\textrm{B}}_{\Delta}(z) = \frac{2\pi}{\sin(\pi \Delta)}\left[z^{-2\Df+2}p_\Delta\!\left(z\right) + \left(\mbox{$\frac{z}{z-1}$}\right)^{-2\Df+2}\!\!p_\Delta\!\left(\mbox{$\frac{z}{z-1}$}\right) + q^{\Df}_{\Delta}(z)\right]\,,
\label{eq:HBGenInt}
\ee
where the correction term $q^{\Df}_{\Delta}(z)$ looks as follows
\be
q^{\Df}_{\Delta}(z) = \frac{a^{\Df}_{\Delta}(z) + b^{\Df}_{\Delta}(z)\log(1-z)}{z^{2\Df-2}}\,.
\label{eq:qGeneral}
\ee
For each $\Df\in\mathbb{N}$, $a^{\Df}_{\Delta}(z)$ and $b^{\Df}_{\Delta}(z)$ are polynomials in $\Delta$ and $z$ which are uniquely fixed by the absence of poles of $H^{\textrm{B}}_{\Delta}(z)$ at $z=0$. For example:
\ba
\Df=1:\quad a^{1}_{\Delta}(z) &= 0\,,\quad b^{1}_{\Delta}(z) = 0\\
\Df=2:\quad a^{2}_{\Delta}(z) &= z^2+2 z-2\,,\quad b^{2}_{\Delta}(z) = 0\\
\Df=3:\quad a^{3}_{\Delta}(z) &= 
z^4+\frac{\left(\Delta ^4-2 \Delta ^3-7 \Delta ^2+8 \Delta +8\right)}{2} z^3-\\
&\quad-\frac{(\Delta -3) (\Delta -2) (\Delta +1) (\Delta +2)}{2} z^2+4 z-2
\,,\quad b^{3}_{\Delta}(z) = 0\,.
\ea
$b^{\Df}_\Delta(z)\neq 0$ for $\Df\geq 4$ despite what the first three cases suggest.

The same ansatz does not work for $\Df\notin\mathbb{Z}$. However, one can make progress by focusing on the Taylor coefficients of $H^{\textrm{B}}_{\Delta}(z)$ around $z=0$. Based on the formulas obtained for $\Df\in\mathbb{N}$, we conjecture the following form of $H^{\textrm{B}}_{\Delta}(z)$ for general $\Df$ and $\Delta$:
\be
H^{\textrm{B}}_{\Delta}(z) =\frac{2^{4(\Df-1)}}{\Gamma (2 \Df-1)^2}\frac{\Gamma \left(\Df-\frac{\Delta}{2}\right)^2 \Gamma \left(\Df-\frac{1-\Delta}{2}\right)^2}{\Gamma \left(1-\frac{\Delta}{2}\right)^2 \Gamma \left(1-\frac{1-\Delta}{2}\right)^2}\frac{2\pi}{\sin (\pi  \Delta)}\sum\limits_{j=0}^{\infty}d^{\textrm{B}}_j(\Df,\Delta)z^{j}\,,
\label{eq:HBGeneral}
\ee
where $d^{\textrm{B}}_{j}(\Df,\Delta)$ are rational functions of $\Df$ and $\Delta$ satisfying $d^{\textrm{B}}_{j}(\Df,\Delta)=d^{\textrm{B}}_{j}(\Df,1-\Delta)$. The first few of them read\footnote{We found the coefficients $d^{\textrm{B}}_{j}(\Df,\Delta)$ for $j=0,\ldots,19$. While they exhibit a lot of structure, we did not find a closed form expression for general $j$. A Mathematica file containing formulas for $d^{\textrm{B}}_{j}(\Df,\Delta)$ and $d^{\textrm{F}}_{j}(\Df,\Delta)$ for higher $j$ is included with the arXiv submission.}
\ba
d^{\textrm{B}}_0(\Df,\Delta) &= 2\\
d^{\textrm{B}}_1(\Df,\Delta) &= 0\\
d^{\textrm{B}}_2(\Df,\Delta) &=
\frac{(\Delta -1)^2 \Delta ^2}{2\Df^2 (2\Df-1)^2}-\frac{2 (\Df-1) (\Delta -1) \Delta }{\Df(2\Df-1)^2}+\frac{(\Df-1)^2 (\Delta -1) \Delta }{(2\Df-1)^2 (\Delta -2) (\Delta +1)}
\,.
\ea

We have performed a number of checks both of the functional equation \eqref{eq:feIH} and the inversion formula \eqref{eq:invPrincipal} using this form of $H^{\textrm{B}}_{\Delta}(z)$ for various values of the parameters (including transcendental $\Df$) which makes us certain of the validity of the above claims. Another very nontrivial check comes from the requirement that for $\Df\notin\mathbb{Z}$ and $\Delta\in2\mathbb{N}$, the inversion kernel should reduce to the partial wave \eqref{eq:phiReduction}. Our result \eqref{eq:HBGeneral} is in complete agreement with this condition.

\subsection{The fermionic case}
The same strategy that worked for the bosonic case and $\Df\in\mathbb{N}$ works for the fermionic case and $\Df\in\mathbb{N}-\frac{1}{2}$. For example, it is easy to check that for $\Df=\frac{1}{2}$, the following expression solves all the constraints
\be
H^{\textrm{F}}_{\Delta}(z) = -\frac{2\pi}{\sin(\pi \Delta)}\left[z\,p_\Delta\!\left(z\right) + \mbox{$\frac{z}{z-1}$}\,p_\Delta\!\left(\mbox{$\frac{z}{z-1}$}\right)\right]
\qquad\left(\Df=1/2\right)\,.
\label{eq:HFHalf}
\ee
As promised, $H^{\textrm{F}}_{\Delta}(z) = O(z^2)$ as $z\rightarrow 0$, guaranteeing the inversion formula works for all physical (and thus Regge-bounded) four-point functions. Analogously to the bosonic case, the inversion kernel for $\Df\in\mathbb{N}-\frac{1}{2}$ takes the general form
\be
H^{\textrm{F}}_{\Delta}(z) = -\frac{2\pi}{\sin(\pi \Delta)}\left[z^{-2\Df+2}p_\Delta\!\left(z\right) + \left(\mbox{$\frac{z}{z-1}$}\right)^{-2\Df+2}\!\!p_\Delta\!\left(\mbox{$\frac{z}{z-1}$}\right) + q^{\Df}_{\Delta}(z)\right]\,,
\label{eq:HFGenInt}
\ee
where $q^{\Df}_{\Delta}(z)$ is given by \eqref{eq:qGeneral}. This time, $a^{\Df}_{\Delta}(z)$ and $b^{\Df}_{\Delta}(z)$ are polynomials in $\Delta$ and $z$ which are fixed by requiring $H^{\textrm{F}}_{\Delta}(z)=O(z^2)$ as $z\rightarrow 0$. The first few cases read
\ba
\Df=1/2:\quad a^{1/2}_{\Delta}(z) &= 0\,,\quad b^{1/2}_{\Delta}(z) = 0\\
\Df=3/2:\quad a^{3/2}_{\Delta}(z) &=\left(2 \Delta ^2-2 \Delta -1\right) z\,,\quad b^{3/2}_{\Delta}(z) = 0\\
\Df=5/2:\quad a^{5/2}_{\Delta}(z) &= \frac{1}{105} \left(2 \Delta ^6-6 \Delta ^5-19 \Delta ^4+48 \Delta ^3+191 \Delta ^2-216 \Delta -243\right)\times\\
&\quad\times z \left(z^2-z+1\right)\,,\\
b^{5/2}_{\Delta}(z) &= \frac{1}{210} \left(2 \Delta ^6-6 \Delta ^5-19 \Delta ^4+48 \Delta ^3-19 \Delta ^2-6 \Delta +72\right)\times\\
&\quad\times(z-2) \left(2 z^2+z-1\right)\,.
\ea
The same general ansatz does not work for $\Df\notin\mathbb{Z}-\frac{1}{2}$. For general $\Df$ and $\Delta$, we conjecture the following Taylor expansion of $H^{\textrm{F}}_{\Delta}(z)$ around $z=0$
\be
H^{\textrm{F}}_{\Delta}(z) =
\frac{2^{2(2\Df-1)}}{\Gamma (2\Df+1)^2}
\frac{\Gamma \left(\Df-\frac{\Delta-1}{2}\right)^2 \Gamma \left(\Df+\frac{\Delta }{2}\right)^2}{\Gamma \left(1-\frac{\Delta }{2}\right)^2 \Gamma \left(\frac{1}{2}+\frac{\Delta }{2}\right)^2}
\frac{2 \pi }{\sin (\pi  \Delta )}
\sum\limits_{j=2}^{\infty}d^{\textrm{F}}_j(\Df,\Delta)z^{j}\,,
\label{eq:HFGeneral}
\ee
where $d^{\textrm{F}}_{j}(\Df,\Delta)$ are rational functions of $\Df$ and $\Delta$ satisfying $d^{\textrm{F}}_{j}(\Df,\Delta)=d^{\textrm{F}}_{j}(\Df,1-\Delta)$. The first two of them read
\be
d^{\textrm{F}}_2(\Df,\Delta)=d^{\textrm{F}}_3(\Df,\Delta) = \frac{(2\Df-1)^2}{2 (\Delta -2) (\Delta +1)}+2 \Delta ^2-2 \Delta +1\,.
\ee

\section{Exchange Witten diagrams}\label{sec:witten}
\subsection{Inverting a single block in the crossed channel}
The main goal of this section is to explain the connection of the presented inversion formula to exchange Witten diagrams.

The existence of the fermionic Lorentzian inversion formula guarantees that a physical (crossing-symmetric and Regge-bounded) four-point function $\mathcal{G}(z)$ is uniquely fixed by its fermionic double discontinuity \eqref{eq:dDiscDef}. Indeed, we can simply insert $\dDisc_{\textrm{F}}\!\left[\mathcal{G}(z)\right]$ into the inversion formulas for the discrete and principal series \eqref{eq:invDiscrete} and \eqref{eq:invPrincipal} to find unique coefficient functions $\widetilde{I}_m$ and $I_{\Delta}$. The four-point function then must be equal to the decomposition \eqref{eq:opeFromIs}. The same comments apply in the bosonic case assuming $z^{-2\Df}\mathcal{G}(z)$ is super-bounded.

The operation of taking the double discontinuity commutes with expanding $\mathcal{G}(z)$ in the t-channel OPE. It is therefore natural to ask which crossing-symmetric, Regge-bounded four-point function (if any) has $\dDisc$ equal to the $\dDisc$ of a single t-channel conformal block of dimension $\Delta$. The short answer is that such function indeed exists and is equal to the symmetric sum of exchange diagrams in $AdS_2$ in the s-, t- and u-channel, where the exchanged field has dimension $\Delta$. In the bosonic case, one also needs to add a uniquely determined contact interaction. 

Focusing on the bosonic case to begin, consider the s-channel $AdS_2$ exchange diagram $W^{(s)}_{\Delta}(z)$ with external scalar propagators of dimension $\Df$ and internal scalar propagator of dimension $\Delta$. Its s-channel OPE contains the single-trace conformal block $G_{\Delta}(z)$, as well as parity-even double-traces $G_{2\Df+2n}(z)$ and their $\Delta$-derivatives $\partial_{\Delta}G_{2\Df+2n}(z)$ for $n=0,1,\ldots$. We normalize $W^{\textrm{(s)}}_{\Delta}(z)$ so that the single-trace $G_{\Delta}(z)$ appears with a unit coefficient. In the u-channel Regge limit $z\rightarrow i\infty$, we have
\be
z^{-2\Df}W^{(s)}_{\Delta}(z) = O(z^{-2})\,,
\ee
so that $z^{-2\Df}W^{(s)}_{\Delta}(z)$ is super-bounded. Here we are taking the limit of the analytic continuation of the function to which $W^{(s)}_{\Delta}(z)$ reduces for $z\in(0,1)$. See \cite{Mazac:2018ycv} for a more detailed discussion of the meaning of the Regge limit in $AdS_2$. 

The t-channel exchange diagram is obtained from crossing
\be
W^{(t)}_{\Delta}(z) = \left|\mbox{$\frac{z}{1-z}$}\right|^{2\Df}W^{(s)}_{\Delta}(1-z)\,.
\ee
The s-channel OPE of $W^{(t)}_{\Delta}(z)$ contains both parity even and odd double-traces $G_{2\Df+j}(z)$ and their $\Delta$-derivatives $\partial_{\Delta}G_{2\Df+j}(z)$ for $j=0,1,\ldots$. By symmetry between the s- and t-channel from the point of view of the u-channel, we have $z^{-2\Df}W^{(t)}_{\Delta}(z) = O(z^{-2})$ as $z\rightarrow i\infty$.

Finally, the u-channel exchange diagram is related to the s-channel diagram by
\be
W^{(u)}_{\Delta}(z) = |z|^{2\Df}W^{(s)}_{\Delta}\left(\mbox{$\frac{1}{z}$}\right)\,.
\ee
The s-channel OPE of $W^{(u)}_{\Delta}(z)$ is identical to the s-channel OPE of $W^{(t)}_{\Delta}(z)$, up to an extra sign $(-1)^j$ for every double-trace of dimension $2\Df+j$. This means that the combination $W^{(t)}_{\Delta}(z)+W^{(u)}_{\Delta}(z)$ only contains the parity-even double traces. This time, we find that in the limit $z\rightarrow i\infty$
\be
z^{-2\Df}W^{(u)}_{\Delta}(z) \sim \delta(\Delta,\Df)z^{-1}\,,
\ee
where $\delta(\Delta,\Df)$ is a computable function. This means $z^{-2\Df}W^{(u)}_{\Delta}(z)$ is not super-bounded but only bounded. In summary, the combination
\be
W^{(s)}_{\Delta}(z)+W^{(t)}_{\Delta}(z)+W^{(u)}_{\Delta}(z)
\ee
is fully crossing-symmetric (with the bosonic statistics). Moreover, the t-channel $\dDisc$ of this function is equal to the $\dDisc$ of a t-channel conformal block of dimension $\Delta$ since $\dDisc$ annihilates all the double-traces of dimension $2\Df+2n$ and their $\Delta$-derivatives. The only problem with this function is that it is not super-bounded, due to the failure of $z^{-2\Df}W^{(u)}_{\Delta}(z)$ to be super-bounded. Luckily, there is a unique possible improvement which is fully crossing-symmetric, has vanishing double discontinuity and makes the function super-bounded. This is precisely the scalar contact diagram of the $\phi^4$ interaction in $AdS_2$, which we denote $A(z)$. Indeed, $A(z)$ only contains the even double-traces and their derivatives in its OPE. Furthermore, we have $z^{-2\Df}A(z)\sim z^{-1}$ as $z\rightarrow i\infty$, which sets the normalization of $A(z)$. Therefore, let us define the following function\footnote{We suppress its $\Df$-dependence to simplify notation.}
\be
P^{\textrm{B}}_{\Delta}(z) = W^{(s)}_{\Delta}(z)+W^{(t)}_{\Delta}(z)+W^{(u)}_{\Delta}(z) - \delta(\Delta,\Df)A(z)\,.
\ee
We will call this function the bosonic Polyakov block for reasons that will become clear in the next section. $P^{\textrm{B}}_{\Delta}(z)$ is the unique Bose-symmetric function such that $z^{-2\Df}P^{\textrm{B}}_{\Delta}(z)$ is super-bounded and such that
\be
\dDisc\!\left[P^{\textrm{B}}_{\Delta}(z)\right] = \dDisc\!\left[G^{(t)}_{\Delta}(z)\right] = 2\sin^2\left[\frac{\pi}{2}(\Delta-2\Df)\right]\left(\mbox{$\frac{z}{1-z}$}\right)^{2\Df}G_{\Delta}(1-z)\,.
\ee
It is an instructive exercise for the reader to convince themselves that no other proposal, such as $G^{(s)}_{\Delta}(z)+G^{(t)}_{\Delta}(z)$, $G^{(s)}_{\Delta}(z)+G^{(t)}_{\Delta}(z)+G^{(u)}_{\Delta}(z)$ or $W^{(s)}_{\Delta}(z)+W^{(t)}_{\Delta}(z)$ satisfies the same requirements.

We can use the Lorentzian inversion formula for the principal series to efficiently find the OPE decomposition of $P_{\Delta}^{\textrm{B}}(z)$. As explained, the s-channel OPE looks as follows
\be
P^{\textrm{B}}_{\Delta}(z) = G_{\Delta}(z) - \sum\limits_{n=0}^{\infty}\left[\alpha^{\textrm{B}}_n(\Delta)G_{\Delta^{\textrm{B}}_n}(z)+\beta^{\textrm{B}}_n(\Delta)\partial G_{\Delta^{\textrm{B}}_n}(z)\right]\,,
\label{eq:pBOPE}
\ee
where $\Delta^{\textrm{B}}_n = 2\Df + 2n$ are the double-trace scaling dimensions and we use the simplified notation $\partial G_{\Delta}(z) \equiv \partial_{\Delta}G_\Delta(z)$. The minus in front of the sum is a useful convention. All the nontrivial structure of exchange diagrams is contained in the coefficients $\alpha^{\textrm{B}}_n(\Delta)$ and $\beta^{\textrm{B}}_n(\Delta)$.

The OPE is encoded in the principal series coefficient function $I_{h}$, where we switch from label $\Delta$ to $h$ to avoid confusion with the scaling dimension labelling the Polyakov block. We will denote the coefficient function for $P^{\textrm{B}}_{\Delta}(z)$ as $\mathcal{I}_{\textrm{B}}(h;\Delta|\Df)$. Terms in the OPE decomposition of $P^{\textrm{B}}_{\Delta}(z)$ translate to poles of $\mathcal{I}_{\textrm{B}}(h;\Delta|\Df)$ in variable $h$. The Lorentzian inversion formula leads to
\be
\mathcal{I}_{\textrm{B}}(h;\Delta|\Df) =
4\sin^2\left[\frac{\pi}{2}(\Delta-2\Df)\right]
\!\!\int\limits_{0}^{1}\!\! dz z^{-2}H^{\textrm{B}}_{h}(z)\left(\mbox{$\frac{z}{1-z}$}\right)^{2\Df}G_{\Delta}(1-z)\,.
\label{eq:iCalB}
\ee
The integral converges for $h$ on the principal series and for $\mathrm{Re}(\Delta)$ sufficiently large (so that $P^{\textrm{B}}_{\Delta}(z)$ is normalizable). For other values of $\Delta$, it can be defined by an analytic continuation in $\Delta$. Let us explain how the expected OPE arises from \eqref{eq:iCalB}. Firstly, recall from \eqref{eq:HBGeneral} that for generic $\Df$, the inversion kernel $H^{\textrm{B}}_{h}(z)$ has double poles in $h$ at the double-trace dimensions $\Delta^{\textrm{B}}_n$, coming from the factor $\Gamma\left(\Df-\frac{h}{2}\right)^2$ in front. These poles give rise to all the double-trace contributions in the OPE \eqref{eq:pBOPE}. To make this more concrete, first note that equation \eqref{eq:opeFromIs} implies the following relationship between the pole structure of $I_{h}$ and the OPE decomposition
\ba
&\frac{I_h}{2K_h}\stackrel{h\rightarrow\widetilde{h}}{\sim}\frac{\alpha}{h-\widetilde{h}}\quad\Leftrightarrow\quad \mathcal{G}(z) =\ldots -\alpha\,G_{\widetilde{h}}(z)+\ldots \\
&\frac{I_h}{2K_h}\stackrel{h\rightarrow\widetilde{h}}{\sim}\frac{\beta}{(h-\widetilde{h})^2}\quad\Leftrightarrow\quad \mathcal{G}(z)=\ldots -\beta\,\partial G_{\widetilde{h}}(z)+\ldots
\ea
Let us therefore define $\widehat{H}^{\textrm{B}}_{n,1}(z)$ and $\widehat{H}^{\textrm{B}}_{n,2}(z)$ as the coefficients of the simple and double pole of the inversion kernels at the double-trace dimensions
\be
\frac{ H^{\textrm{B}}_{h}(z)}{K_{h}} \stackrel{h\rightarrow\Delta^{\textrm{B}}_n}{=} \frac{\widehat{H}^{\textrm{B}}_{n,2}(z)}{(h-\Delta^{\textrm{B}}_n)^2} + \frac{\widehat{H}^{\textrm{B}}_{n,1}(z)}{h-\Delta^{\textrm{B}}_n} + \textrm{finite}\,.
\label{eq:hHatBDef}
\ee
Plugging this expression into the inversion integral \eqref{eq:iCalB} leads to the following formulas for the coefficients $\alpha^{\textrm{B}}_n(\Delta)$ and $\beta^{\textrm{B}}_n(\Delta)$
\ba
\alpha^{\textrm{B}}_n(\Delta) &=
2\sin^2\left[\frac{\pi}{2}(\Delta-2\Df)\right]
\!\!\int\limits_{0}^{1}\!\! dz z^{-2}\,\widehat{H}^{\textrm{B}}_{n,1}(z)\left(\mbox{$\frac{z}{1-z}$}\right)^{2\Df}G_{\Delta}(1-z)\\
\beta^{\textrm{B}}_n(\Delta) &=
2\sin^2\left[\frac{\pi}{2}(\Delta-2\Df)\right]
\!\!\int\limits_{0}^{1}\!\! dz z^{-2}\,\widehat{H}^{\textrm{B}}_{n,2}(z)\left(\mbox{$\frac{z}{1-z}$}\right)^{2\Df}G_{\Delta}(1-z)\,.
\label{eq:abB}
\ea
In the special case $\Df\in\mathbb{N}$, $\alpha^{\textrm{B}}_n(\Delta)$ will also receive a contribution from the discrete series sum. In effect, this means that in the above equation for $\alpha^{\textrm{B}}_n(\Delta)$, we should make the replacement
\be
\widehat{H}^{\textrm{B}}_{n,1}(z) \mapsto\widehat{H}^{\textrm{B}}_{n,1}(z) - \frac{2\Gamma(\Delta^{\textrm{B}}_n)^4}{\pi^2\Gamma(2\Delta^{\textrm{B}}_n)\Gamma(2\Delta^{\textrm{B}}_n-1)}G_{\Delta^{\textrm{B}}_n}(z)\qquad\textrm{for }\Df\in\mathbb{N}.
\ee

Having understood how the double-trace contributions to the OPE of $P^{\textrm{B}}_{\Delta}(z)$ arise from the point of view of the Lorentzian inversion formula, it remains to be seen that the single-trace conformal block $G_{\Delta}(z)$ is also present. The corresponding pole of $\mathcal{I}_{\textrm{B}}(h;\Delta|\Df)$ can not come from a pole of $H^{\textrm{B}}_{h}(z)$ since its location depends on $\Delta$. Instead it must come from a singularity of the integral over $z$ as $z\rightarrow 1$ according to
\be
\int\limits_0^1\!\!dz\,(1-z)^{\Delta-h-1} = -\frac{1}{h-\Delta}\,.
\ee
In fact, in all the cases where $ H^{\textrm{B}}_{h}(z)$ is under control, we found that its leading term in the $z\rightarrow 1$ expansion takes the form
\be
 H^{\textrm{B}}_{h}(z) \stackrel{z\rightarrow1}{\sim} \frac{K_{h}}{2\sin^2\left[\frac{\pi}{2}(h-2\Df)\right]}(1-z)^{2\Df-h-1}\,.
 \label{eq:z1Power}
\ee
Plugging this expansion into the inversion formula \eqref{eq:iCalB} and using the $z\rightarrow 1$ expansion of the t-channel conformal block, we find
\be
\frac{\mathcal{I}_{\textrm{B}}(h;\Delta|\Df)}{2K_{h}} \stackrel{h\rightarrow\Delta}{\sim} - \frac{1}{h-\Delta}\,,
\ee
thus precisely reproducing the single-trace term in $P^{\textrm{B}}_{\Delta}(z)$.

\subsection{Fermionic Polyakov blocks}
It is relatively straightforward to adapt the discussion of the previous subsection to the fermionic case. The task is to find a fully Fermi-symmetric four-point function, i.e. one satisfying \eqref{eq:extensionF} whose $\dDisc$ agrees with $\dDisc_{\textrm{F}}$ of a t-channel block of dimension $\Delta$
\be
\dDisc_{\textrm{F}}\!\left[ G^{(t)}_{\Delta}(z)\right]=2\cos^2\!\left[\frac{\pi}{2}(\Delta-2\Df)\right]\left(\mbox{$\frac{z}{1-z}$}\right)^{2\Df}G_{\Delta}(1-z)\,.
\ee
The sought-after function must also be bounded (but not necessarily super-bounded) in the Regge limit. The answer is given in terms of exchange Witten diagrams with external fermionic propagators of dimension $\Df$ and internal scalar propagator of dimension $\Delta$. We will denote the s-channel diagram as $V^{(s)}_{\Delta}(z)$. In fact, $V^{(s)}_{\Delta}(z)$ is equal to the standard bosonic exchange diagram with the external dimension shifted by $+1/2$, see \cite{Faller:2017hyt}
\be
V^{(s)}_{\Delta,\Df}(z) = W^{(s)}_{\Delta,\Df+\frac{1}{2}}(z)\,.
\ee
This relation also makes it clear that the s-channel OPE of $V^{(s)}_{\Delta}(z)$ contains the single-trace block $G_{\Delta}(z)$ (with coefficient one) as well as the fermionic double-traces of dimensions $\Delta^{\textrm{F}}_n = 2\Df+2n+1$, $n=0,1,\ldots$. It also follows from the above relation and the known Regge behaviour of bosonic exchanges that $z^{-2\Df}V^{(s)}_{\Delta}(z)=O(z^{-1})$ as $z\rightarrow i\infty$, which is therefore Regge-bounded. The t- and u-channel diagrams are obtained from the s-channel diagram by crossing transformations
\ba
V^{(t)}_{\Delta,\Df}(z) &= \sgn\left(\mbox{$\frac{z}{1-z}$}\right)\left|\mbox{$\frac{z}{1-z}$}\right|^{2\Df}V^{(s)}_{\Delta,\Df}(1-z) = \left(\mbox{$\frac{z}{1-z}$}\right)^{-1}W^{(t)}_{\Delta,\Df+\frac{1}{2}}(z)\\
V^{(u)}_{\Delta,\Df}(z) &= -\sgn\left(z\right)\left|z\right|^{2\Df}V^{(s)}_{\Delta,\Df} \left(\mbox{$\frac{1}{z}$}\right) = -z^{-1}W^{(u)}_{\Delta,\Df+\frac{1}{2}}(z)\,.
\ea
In particular $z^{-2\Df}V^{(t)}_{\Delta}(z) = O(z^{-1})$ and $z^{-2\Df}V^{(u)}_{\Delta}(z) = O(z^{-1})$ as $z\rightarrow i\infty$ and so are both bounded there.

These results make it clear that the sought-after function is
\be
P^{\textrm{F}}_{\Delta}(z) = V^{(s)}_{\Delta}(z)+V^{(t)}_{\Delta}(z)+V^{(u)}_{\Delta}(z)\,.
\label{eq:pF}
\ee
We will call it the fermionic Polyakov block. No contact diagram correction is needed since $P^{\textrm{F}}_{\Delta}(z)$ is automatically Regge-bounded. In fact, no Regge-bounded contact diagram exists since the simplest non-vanishing bulk vertex with four fermions has two derivatives. The OPE decomposition of the fermionic Polyakov blocks takes the form
\be
P^{\textrm{F}}_{\Delta}(z) = G_{\Delta}(z) - \sum\limits_{n=0}^{\infty}\left[\alpha^{\textrm{F}}_n(\Delta)G_{\Delta^{\textrm{F}}_n}(z)+\beta^{\textrm{F}}_n(\Delta)\partial G_{\Delta^{\textrm{F}}_n}(z)\right]\,.
\label{eq:pFOPE}
\ee
The principal series coefficient function $I_h$ of $P^{\textrm{F}}_{\Delta}(z)$ will be denoted $\mathcal{I}_{\textrm{F}}(h;\Delta|\Df)$. It can be computed using the fermionic version of the Lorentzian inversion formula as follows
\be
\mathcal{I}_{\textrm{F}}(h;\Delta|\Df) =
4\cos^2\left[\frac{\pi}{2}(\Delta-2\Df)\right]
\!\!\int\limits_{0}^{1}\!\! dz z^{-2}H^{\textrm{F}}_{h}(z)\left(\mbox{$\frac{z}{1-z}$}\right)^{2\Df}G_{\Delta}(1-z)\,.
\label{eq:iCalF}
\ee
In order to compute the OPE coefficients $\alpha^{\textrm{F}}_n(\Delta)$ and $\beta^{\textrm{F}}_n(\Delta)$, let us define the residues of the fermionic inversion kernel $\widehat{H}^{\textrm{F}}_{n,1}(z)$, $\widehat{H}^{\textrm{F}}_{n,2}(z)$ as follows
\be
\frac{ H^{\textrm{F}}_{h}(z)}{K_{h}} \stackrel{h\rightarrow\Delta^{\textrm{F}}_n}{=} \frac{\widehat{H}^{\textrm{F}}_{n,2}(z)}{(h-\Delta^{\textrm{F}}_n)^2} + \frac{\widehat{H}^{\textrm{F}}_{n,1}(z)}{h-\Delta^{\textrm{F}}_n} +\textrm{finite}\,.
\label{eq:hHatFDef}
\ee
The OPE coefficients then read
\ba
\alpha^{\textrm{F}}_n(\Delta) &=
2\cos^2\left[\frac{\pi}{2}(\Delta-2\Df)\right]
\!\!\int\limits_{0}^{1}\!\! dz z^{-2}\,\widehat{H}^{\textrm{F}}_{n,1}(z)\left(\mbox{$\frac{z}{1-z}$}\right)^{2\Df}G_{\Delta}(1-z)\\
\beta^{\textrm{F}}_n(\Delta) &=
2\cos^2\left[\frac{\pi}{2}(\Delta-2\Df)\right]
\!\!\int\limits_{0}^{1}\!\! dz z^{-2}\,\widehat{H}^{\textrm{F}}_{n,2}(z)\left(\mbox{$\frac{z}{1-z}$}\right)^{2\Df}G_{\Delta}(1-z)\,.
\label{eq:abF}
\ea
In the special case $\Df\in\mathbb{N}-\frac{1}{2}$, $\alpha^{\textrm{F}}_n(\Delta)$ receives an extra contribution from the discrete series, meaning we should make the following replacement in the above formula
\be
\widehat{H}^{\textrm{F}}_{n,1}(z) \mapsto\widehat{H}^{\textrm{F}}_{n,1}(z) - \frac{2\Gamma(\Delta^{\textrm{F}}_n)^4}{\pi^2\Gamma(2\Delta^{\textrm{F}}_n)\Gamma(2\Delta^{\textrm{F}}_n-1)}G_{\Delta^{\textrm{F}}_n}(z)\qquad\textrm{for }\Df\in\mathbb{N}-\frac{1}{2}.
\ee

\subsection{Explicit results for the coefficient function}
The Lorentzian inversion integrals \eqref{eq:iCalB} and \eqref{eq:iCalF} for $\mathcal{I}_{\textrm{B}}(h;\Delta|\Df)$ and $\mathcal{I}_{\textrm{F}}(h;\Delta|\Df)$ can be evaluated explicitly in the cases where the full inversion kernel is known, i.e. for $\Df\in\mathbb{N}$ in the bosonic case and $\Df\in\mathbb{N}-\frac{1}{2}$ in the fermionic case.  Firstly, recall from \eqref{eq:HBGenInt} and \eqref{eq:HFGenInt} that both $H^{\textrm{B}}_{\Delta}(z)$ and $H^{\textrm{F}}_{\Delta}(z)$ always contain the term
\be
 \frac{2\pi}{\sin(\pi \Delta)}\left(\mbox{$\frac{z}{1-z}$}\right)^{-2\Df+2}\!\!p_\Delta\!\left(\mbox{$\frac{z}{z-1}$}\right)\,,
 \ee
leading to the following term in $\mathcal{I}_{\textrm{B,F}}(h;\Delta|\Df)$ (for the above discrete values of $\Df$)
\ba
\mathcal{I}_{\textrm{B,F}}(h;\Delta|\Df)&\supset\frac{8\pi\sin^2\left(\frac{\pi\Delta}{2}\right)}{\sin(\pi h)}
\!\!\int\limits_{0}^{1}\!\! dz z^{-2}p_h\!\left(\mbox{$\frac{z-1}{z}$}\right)G_{\Delta}(z) =\\
&=-\frac{8\pi\sin^2\left(\frac{\pi\Delta}{2}\right)}{\sin(\pi h)}\frac{\Gamma(2\Delta)}{\Gamma(\Delta)^2}
\frac{1}{(h-\Delta)(h+\Delta-1)}\,.
\label{eq:ICalSPart}
\ea
The integral gives a simple answer since $p_\Delta\!\left(\mbox{$\frac{z-1}{z}$}\right)$ and $G_{\Delta}(z)$ are both eigenfunctions of the s-channel Casimir. This term reproduces correctly the single-trace pole at $h=\Delta$ and is very similar to the coefficient function of the full s-channel exchange diagram $W^{(s)}_{\Delta}(z)$. The latter takes the following form for general (i.e. not only integer) $\Df$
\be
\mathcal{I}^{(s)}_{\textrm{B}}(h;\Delta|\Df) =
-\frac{\left(\frac{h+1}{2}\right)_{\Df-1}^2\left(1-\frac{h}{2}\right)_{\Df-1}^2}{\left(\frac{\Delta +1}{2}\right)_{\Df-1}^2 \left(1-\frac{\Delta }{2}\right)_{\Df-1}^2}\frac{8\pi\sin^2\left(\frac{\pi\Delta}{2}\right)}{\sin(\pi h)}\frac{\Gamma(2\Delta)}{\Gamma(\Delta)^2}
\frac{1}{(h-\Delta)(h+\Delta-1)}\,,
\ee
where $(x)_y=\frac{\Gamma(x+y)}{\Gamma(x)}$, i.e. it only differs from the expression in \eqref{eq:ICalSPart} by the ratio of Pochhammer symbols in front. The prefactor is necessary to remove poles at $h=2\Df-2,2\Df-4,\ldots$. As stated in the previous subsection, the fermionic s-channel Witten diagram $V^{(s)}_{\Delta}(z)$ is obtained from $W^{(s)}_{\Delta}(z)$ by a shift in $\Df$
\be
\mathcal{I}^{(s)}_{\textrm{F}}(h;\Delta|\Df) = \mathcal{I}^{(s)}_{\textrm{B}}(h;\Delta|\Df+1/2)\,,
\ee
meaning \eqref{eq:ICalSPart} precisely agrees with the fermionic s-channel diagram for $\Df=\frac{1}{2}$.

The remaining contributions to \eqref{eq:HBGenInt} and \eqref{eq:HFGenInt} must reproduce the rest of the Polyakov blocks, i.e. the coefficient functions of the crossed-channel exchange diagrams (together with a contact term in the bosonic case), as well as the difference between the s-channel diagram and \eqref{eq:ICalSPart}. The contribution of the first term in the square brackets in \eqref{eq:HBGenInt} and \eqref{eq:HFGenInt} can be evaluated using the general formula
\be
\int\limits_{0}^{1}\!\!\d z\,p_{h}(z)\frac{G_{\Delta}(1-z)}{(1-z)^{2\Df}} = \frac{\Gamma(2\Delta)}{\Gamma(\Delta)^2}s(h;\Delta|\Df)
\ee
where
\ba
s(h;\Delta|\Df) = & \frac{\Gamma(\Delta)^2\Gamma(\Delta-2\Df+1)^2}{\Gamma(2\Delta)\Gamma(\Delta-h-2\Df+2)\Gamma(\Delta+h-2\Df+1)}\times\\
&\times{}_4F_3\left( {\begin{array}{*{20}{c}}
{\Delta ,\Delta ,\Delta -2 \Df+1,\Delta -2 \Df+1}\\
{2 \Delta,\Delta-h -2 \Df+2 ,\Delta+h -2 \Df+1}
\end{array};1} \right)\,.
\ea
Putting the contributions together, we find
\ba
\mathcal{I}_{\textrm{B}}(h;\Delta|\Df) = &\frac{8\pi\sin^2\left(\frac{\pi\Delta}{2}\right)}{\sin(\pi h)}\frac{\Gamma(2\Delta)}{\Gamma(\Delta)^2}\left[-\frac{1}{(h-\Delta)(h+\Delta-1)}+\right.\\
&\qquad\qquad\qquad\qquad\left.\phantom{\frac{1}{1}}+s(h;\Delta|\Df)+r_{\textrm{B}}(h;\Delta|\Df) \right]\quad
\textrm{for }\Df\in\mathbb{N}
\label{eq:ICalBGen}
\ea
and
\ba
\mathcal{I}_{\textrm{F}}(h;\Delta|\Df) = &\frac{8\pi\sin^2\left(\frac{\pi\Delta}{2}\right)}{\sin(\pi h)}\frac{\Gamma(2\Delta)}{\Gamma(\Delta)^2}\left[-\frac{1}{(h-\Delta)(h+\Delta-1)}-\right.\\
&\qquad\qquad\qquad\left.\phantom{\frac{1}{1}}-s(h;\Delta|\Df)-r_{\textrm{F}}(h;\Delta|\Df) \right]\quad
\textrm{for }\Df\in\mathbb{N}-\frac{1}{2}\,.
\label{eq:ICalFGen}
\ea
Here $r_{\textrm{B,F}}(h;\Delta|\Df)$ are the contributions of the last term in the square bracket in \eqref{eq:HBGenInt} and \eqref{eq:HFGenInt}. $r_{\textrm{B,F}}(h;\Delta|\Df)$ are polynomials in $h$ and rational functions in $\Delta$ whose complexity increases with increasing $\Df$. For example
\ba
&r_{\textrm{B}}(h;\Delta|1) = 0\\
&r_{\textrm{B}}(h;\Delta|2) = -\frac{2 \left(\Delta^4-2 \Delta^3-6 \Delta^2+7 \Delta+4\right)}{\left(\Delta-3\right) \left(\Delta-2\right) \left(\Delta-1\right) \Delta \left(\Delta+1\right) \left(\Delta+2\right)}\\
&r_{\textrm{F}}(h;\Delta|1/2) = 0\\
&r_{\textrm{F}}(h;\Delta|3/2) = \frac{2 h^2-2 h-1}{(\Delta -2) (\Delta -1) \Delta  (\Delta +1)}\,.
\ea
Formulas for $r_{\textrm{B,F}}(h;\Delta|\Df)$ for higher values of $\Df$ are included in an attached Mathematica notebook. We have verified that the OPE decomposition following from \eqref{eq:ICalBGen} and \eqref{eq:ICalFGen} agrees with that of the crossing-symmetrized exchange diagrams in $AdS_2$.\footnote{We thank Xinan Zhou for providing us with a draft of \cite{Zhou:2018sfz} to facilitate some of the checks.}

In the future, it would be useful to find closed formulas for the coefficient functions $\mathcal{I}_{\textrm{B,F}}(h;\Delta|\Df)$ for general $\Df$. Optimistically, they should be expressible in terms of generalized hypergeometric functions ${}_{p}F_{p-1}(\ldots;1)$ with $p=4$ or higher. This is what happens for the crossed-channel Witten diagrams in $d=2,4$, as shown in \cite{Liu:2018jhs}. Furthermore, the residues of these coefficient functions at the double-traces appear to be expressible as generalized hypergeometric functions in general spacetime dimension \cite{Sleight:2018ryu,Gopakumar:2018xqi}.

\section{Polyakov bootstrap and extremal functionals}\label{sec:polyakov}
\subsection{Polyakov's approach to the conformal bootstrap}
In the usual approach to the conformal bootstrap, one starts from the existence of OPE which contains only physical operators,\footnote{i.e. only elements of the Hilbert space of the theory on $S^{d-1}$} and imposes crossing symmetry as a constraint on the CFT data. There exists an alternative approach going back to Polyakov's work \cite{Polyakov:1974gs} and recently revived in \cite{Sen:2015doa,Gopakumar2017,Gopakumar2017a}, where in some sense the reverse is done. Specifically, one postulates the existence of universal (i.e. theory-independent) functions $P_{\Delta,J}(z,\bar{z})$ which are crossing-symmetric
\be
P_{\Delta,J}(z,\bar{z}) = \left[\mbox{$\frac{z\bar{z}}{(1-z)(1-\bar{z})}$}\right]^{\Df}P_{\Delta,J}(1-z,1-\bar{z}) =
\left(z \bar{z} \right)^{\Df}P_{\Delta,J}\left(\mbox{$\frac{1}{z},\frac{1}{\bar{z}}$}\right)\,,
\ee
and which have the property that if one takes the OPE expansion of a four-point function (of identical scalars for simplicity), and replaces the conformal blocks $G_{\Delta,J}(z,\bar{z})$ by $P_{\Delta,J}(z,\bar{z})$, one gets the same answer
\be
\mathcal{G}(z,\bar{z}) = \sum\limits_{\mathcal{O}\in\phi\times\phi}(c_{\phi\phi\mathcal{O}})^2G_{\Delta_{\mathcal{O}},J_{\mathcal{O}}}(z,\bar{z})\stackrel{?}{=}
\sum\limits_{\mathcal{O}\in\phi\times\phi}(c_{\phi\phi\mathcal{O}})^2P_{\Delta_{\mathcal{O}},J_{\mathcal{O}}}(z,\bar{z})\,.
\label{eq:polyakovBoo}
\ee
It is not a priori clear whether such functions $P_{\Delta,J}(z,\bar{z})$ indeed exist. We will refer to these hypothetical functions as the Polyakov blocks. It has been proposed that $P_{\Delta,J}(z,\bar{z})$ exist and are equal to the sum of the s-, t- and u-channel Witten diagrams, supplemented by appropriate contact terms \cite{Gopakumar2017,Gopakumar2017a,Dey:2017fab}. However, no universal prescription for fixing the contact terms or a general proof of consistency of this approach has been presented. The OPE expansion of $P_{\Delta,J}(z,\bar{z})$ then contains the single-trace conformal block $G_{\Delta,J}(z,\bar{z})$ together with double-trace contributions. The conformal bootstrap constraints are precisely the requirement that these unphysical double-trace contributions cancel out after performing the sum over physical operators in \eqref{eq:polyakovBoo}. Recent literature on the Polyakov bootstrap has used Mellin-space techniques to extract the bootstrap equations. However, the meaning of the Polyakov bootstrap can be stated without making reference to Mellin space, as we have just done.

Despite several successful (perturbative) applications of the Polyakov bootstrap in recent literature, its status remains somewhat unclear. In particular, one can ask the following questions
\begin{enumerate}
\item Do Polyakov blocks $P_{\Delta,J}(z,\bar{z})$ with the above properties really exist?
\item If so, how does one fix the contact-term ambiguity?
\item What is the relationship of the Polyakov bootstrap to the standard bootstrap, where physical OPE is manifest and crossing serves as the constraint?
\item Does the approach make sense non-perturbatively or is it inherently perturbative?
\end{enumerate}
These questions were answered in companion work \cite{Mazac:2018ycv} in the $sl(2,\mathbb{R})$ setting using the framework of analytic bootstrap functionals. In the following, we will answer them also in the $sl(2,\mathbb{R})$ setting using the closely-related inversion formula of this note. We will explain the connection between the inversion formula and the functional bootstrap along the way.

The short answers to the above questions are that the $sl(2,\mathbb{R})$ Polyakov blocks exist and are equal to the Polyakov blocks discussed in the previous section, i.e. to the Lorentzian inverse of individual t-channel conformal blocks. This means the inversion formula fixes the contact terms. The resulting bootstrap equations are valid non-perturbatively and are equivalent to the standard crossing equation. We will now present a detailed proof of these claims. The basic idea is that inserting the t-channel OPE into the Lorentzian inversion formula provides the required expansion into Polyakov blocks.

\subsection{The argument}\label{ssec:polyakovArgument}
We work in the fermionic setting for simplicity since then the inversion formula \eqref{eq:invPrincipal} applies to all unitary four-point functions, and not just those super-bounded in the Regge limit as in the bosonic case. This will be amended in \ref{ssec:reggeRelax} when we correct the bosonic inversion formula so that it applies to all unitary four-point functions. Since all objects of this and the following subsection will pertain to the fermionic case, we drop the F sub-/super-scripts to simplify notation.

As in the rest of this note, suppose that $\mathcal{G}(z)$ is a four-point function of identical $sl(2,\mathbb{R})$ primaries in a unitary theory. It is given for $z\in(0,1)$, where it has the following OPE
\be
\mathcal{G}(z) = \!\!\sum\limits_{\mathcal{O}\in\phi\times\phi}\!\!(c_{\phi\phi\mathcal{O}})^2 G_{\Delta_{\mathcal{O}}}(z)\,.
\label{eq:gCalOPE}
\ee
Let us define $\mathcal{G}(z)$ for all $z\in\mathbb{R}$ using the fermionic extension \eqref{eq:extensionF}.

In order for the coefficient function $I_h$ to be well-defined, $\mathcal{G}(z)$ should be normalizable with respect to the scalar product \eqref{eq:innerproduct}. Physical four-point functions are often not normalizable due to the presence of low-lying operators in the OPE. It is easy to check that normalizability at $z=0$ requires $\Delta_{\mathcal{O}}>\frac{1}{2}$ and normalizability at $z=1$ requires $\Delta_{\mathcal{O}}>2\Df-\frac{1}{2}$ for all $\mathcal{O}\in\phi\times\phi$. In particular, the identity operator makes $\mathcal{G}(z)$ non-normalizable. Appendix B.2 of reference \cite{Simmons-Duffin:2017nub} explains how to deal with non-normalizable contributions in the $D>1$ inversion formula. However, their regularization breaks crossing symmetry and thus is not useful for us. Instead, we can regulate the four-point function by subtracting the (finitely many) fermionic Polyakov blocks of the non-normalizable operators. Let us define
\be
\mathcal{G}^{\textrm{reg}}(z)\equiv
\mathcal{G}(z) -\!\!\!\!\!\!\sum\limits_{0\leq\Delta_{\mathcal{O}}\leq\Delta^*}\!\!\!\!\!\!
(c_{\phi\phi\mathcal{O}})^2 P_{\Delta_{\mathcal{O}}}(z)\,,
\ee
where $\Delta^*=\max(\frac{1}{2},2\Df-\frac{1}{2})$. It is not hard to see that the fermionic Polyakov block of the identity operator is simply the mean-field theory four-point function
\be
P_{0}(z) =
1 -\sgn\left(\mbox{$\frac{z}{z-1}$}\right) \left|\mbox{$\frac{z}{z-1}$}\right|^{2\Df} -\sgn(z) |z|^{2\Df}\,.
\ee
$\mathcal{G}^{\textrm{reg}}(z)$ is crossing-symmetric, Regge-bounded and normalizable.\footnote{In fact, normalizability is only guaranteed for $\Df>\frac{1}{4}$ since otherwise the double-trace contributions of the subtracted Polyakov block are themselves non-normalizable at $z=0$. Although we believe all the important conclusions of this section hold also for $\Df\in(0,\frac{1}{4}]$, our argument only applies for $\Df>\frac{1}{4}$ for the above reason.}

We can use the Euclidean inversion formula to compute the coefficient function of $\mathcal{G}^{\textrm{reg}}(z)$
\be
I_h^{\textrm{reg}} = 
\!\!\! \int\limits_{-\infty}^{\infty}\!\!\!dz z^{-2}\,\Psi_{h}(z)\,\mathcal{G}^{\textrm{reg}}(z)\,.
\label{eq:eucInv2}
\ee
The integral converges to a meromorphic function of $h$ in the strip centered at the principal series $1-\Delta_0<\mathrm{Re}(h)<\Delta_0$, where $\Delta_0$ is the leading scaling dimension present in the OPE of $\mathcal{G}^{\textrm{reg}}(z)$.\footnote{The only possible poles in this strip are the poles of $\Psi_h(z)$ at $h=1,3,\ldots$ and their shadow poles. The former cancel in the combination $I_h/K_h$ which gives rise to the OPE.\label{fn:poles}} In order to find the analytic continuation of $I_h^{\textrm{reg}}$ to general complex $h$, we need to start instead from the Lorentzian inversion formula
\be
I^{\textrm{reg}}_{h} = 2\!\!\int\limits_{0}^{1}\!\! dz z^{-2}H_{h}(z)\dDisc\!\left[\mathcal{G}^{\textrm{reg}}(z)\right].
\label{eq:lorInv2}
\ee
The double discontinuity can be expanded using the t-channel OPE
\be
\dDisc\!\left[\mathcal{G}^{\textrm{reg}}(z)\right]=
2\!\!\sum\limits_{\Delta_{\mathcal{O}}>\Delta^*}\!
(c_{\phi\phi\mathcal{O}})^2
\cos^2\!\left[\frac{\pi}{2}(\Delta_{\mathcal{O}}-2\Df)\right]\left(\mbox{$\frac{z}{1-z}$}\right)^{2\Df}G_{\Delta_{\mathcal{O}}}(1-z)\,.
\label{eq:dDiscTChannel}
\ee
Note that the integral \eqref{eq:lorInv2} never has a divergence coming from the $z=0$ endpoint. This is because $H_h(z) = O(z^2)$ and $\dDisc\!\left[\mathcal{G}^{\textrm{reg}}(z)\right] \leq \dDisc\!\left[\mathcal{G}(z)\right]\leq \mathcal{G}(z)=O(z^0)$. On the other hand, the integral diverges at the $z=1$ unless $1-\Delta_0<\mathrm{Re}(h)<\Delta_0$ because of the power-law singularity of $H_{h}(z)$ there, see \eqref{eq:z1Power}.

In other words, the  Euclidean and Lorentzian integrals \eqref{eq:eucInv2} and \eqref{eq:lorInv2} converge in the same region of complex $h$. The advantage of the Lorentzian integral is that it can be expanded using the t-channel OPE. Indeed, the dominated convergence theorem guarantees that the Lorentzian integral can be commuted with the t-channel OPE \eqref{eq:dDiscTChannel} whenever the integral converges. For the theorem to apply, it is crucial that \eqref{eq:dDiscTChannel} gives $\dDisc\!\left[\mathcal{G}^{\textrm{reg}}(z)\right]$ as a sum of \emph{positive} terms. Commuting the sum and integral gives
\be
I^{\textrm{reg}}_h = \!\!\sum\limits_{\Delta_{\mathcal{O}}>\Delta^*}\!
(c_{\phi\phi\mathcal{O}})^2\,\mathcal{I}(h;\Delta_{\mathcal{O}}|\Df)\,,
\ee
where $\mathcal{I}(h;\Delta_{\mathcal{O}}|\Df)$ is the coefficient function of the fermionic Polyakov block of exchanged dimension $\Delta_{\mathcal{O}}$, discussed in the previous section and given by \eqref{eq:iCalF}. We can now add back the Polyakov blocks that we subtracted to make $\mathcal{G}(z)$ normalizable. We find $\mathcal{G}(z)$ is described by the coefficient function
\be
I_h = \!\!\sum\limits_{\mathcal{O}\in\phi\times\phi}\!
(c_{\phi\phi\mathcal{O}})^2\,\mathcal{I}(h;\Delta_{\mathcal{O}}|\Df)\,,
\label{eq:IFOPE}
\ee
where the sum runs over all primary operators in the OPE. As a function of $h$, $\mathcal{I}(h;\Delta|\Df)$ is meromorphic with a simple pole at the single-trace dimension $h=\Delta$, double poles at the double-trace dimensions $h=\Delta_{n} = 2\Df+2n+1$, as well as all the shadow poles (there are also the unphysical poles of $\Psi_h(z)$ described in footnote \ref{fn:poles}).

The argument so far shows that the sum \eqref{eq:IFOPE} converges in the region $1-\Delta_0<\mathrm{Re}(h)<\Delta_0$ away from poles of the individual terms to a meromorphic function of $h$. Let $S$ be the union of the locations of poles in $h$ of all terms appearing in the sum \eqref{eq:IFOPE}. $S$ is a subset of the real axis with no accumulation point (besides $h=\infty$). We will now proceed to show that \eqref{eq:IFOPE} in fact converges for all $h\in\mathbb{C}\backslash S$. For that, we need to understand the behaviour of $\mathcal{I}(h;\Delta|\Df)$ for large real $\Delta$ and general $h\in\mathbb{C}$. In this regime, we can use the convergent integral representation \eqref{eq:iCalF}, which we reproduce here for convenience
\be
\mathcal{I}(h;\Delta|\Df) =
4\cos^2\left[\frac{\pi}{2}(\Delta-2\Df)\right]
\!\!\int\limits_{0}^{1}\!\! dz z^{-2}H_{h}(z)\left(\mbox{$\frac{z}{1-z}$}\right)^{2\Df}G_{\Delta}(1-z)\,.
\ee
As $\Delta\rightarrow+\infty$ along the real axis, the integral is dominated by the region $z\ll 1$. We have seen that $H_h(z) \sim a(\Df,h)z^{2}$ as $z\rightarrow 0$, where $a(\Df,h)$ is a known function holomorphic for $\mathbb{C}\backslash S$. The integral can be evaluated using the saddle-point method, giving
\be
\mathcal{I}(h;\Delta|\Df) \stackrel{\Delta\rightarrow +\infty}{\sim}
a(\Df,h)\frac{2\Gamma (2\Df+1)^2}{\sqrt{\pi }}
\cos^2\left[\frac{\pi}{2}(\Delta-2\Df)\right]2^{2 \Delta} \Delta ^{-4\Df-\frac{3}{2}}\,.
\label{eq:ICalFLargeDelta}
\ee
Crucially, the dependence on $h$ completely factorizes as $\Delta\rightarrow\infty$ and enters only through the prefactor $a(\Df,h)$.

We already know the sum \eqref{eq:IFOPE} converges for $|\mathrm{Re}(h)-\frac{1}{2}|<\epsilon$, but now we see that changing $h$ to a general complex value will not alter the convergence of the tail of the sum, since the $h$-dependence entirely factorizes from the leading $\Delta$-dependence for $\Delta\gg1$. Therefore, \eqref{eq:IFOPE} converges for all $h\in\mathbb{C}\backslash S$. In fact, the upper bound on the OPE data proved in \cite{Mazac:2018ycv} implies that the tail of the sum \eqref{eq:IFOPE} for a general unitary solution to crossing is bounded by its value in the generalized free boson (times a constant). In the later case we have
\be
(c_{\phi\phi\mathcal{O}})^2 \sim
\frac{8\sqrt{\pi }}{\Gamma (2\Df)^2}2^{-2 \Delta_{\mathcal{O}}} \Delta_{\mathcal{O}}^{4\Df-\frac{3}{2}}
\ee
as $\Delta_{\mathcal{O}}\rightarrow\infty$. Together with \eqref{eq:ICalFLargeDelta}, this implies that at large $\Delta$, the contribution of operators in \eqref{eq:IFOPE} with $\Delta_{\mathcal{O}}\in[\Delta-1,\Delta+1]$ is bounded by $\cos^2\left[\frac{\pi}{2}(\Delta-2\Df)\right]\Delta^{-3}$ times an $h$-dependent constant. This in turn implies that the error one would make by truncating the sum \eqref{eq:IFOPE} to operators with $\Delta_{\mathcal{O}}\leq\Delta_{\textrm{max}}$ is always $O(\Delta^{-2}_{\textrm{max}})$.

It follows immediately that \eqref{eq:IFOPE} converges \emph{uniformly} in $h$ on any compact subset of $\mathbb{C}\backslash S$. Since a uniform limit of holomorphic functions is holomorphic, we conclude that \eqref{eq:IFOPE} gives $I_h$ as a function holomorphic in $\mathbb{C}\backslash S$.

\subsection{Sum rules}\label{ssec:sumrules}
The argument of the previous subsection showed that the coefficient function $I_h$ can be written as a convergent sum of coefficient functions of the fermionic Polyakov blocks as in \eqref{eq:IFOPE}. All the poles of $I_h$ come entirely from poles of the individual terms $\mathcal{I}(h;\Delta_{\mathcal{O}}|\Df)$. Furthermore, the residue of $I_h$ at a pole $h_0$ is equal to the sum of residues of $\mathcal{I}(h;\Delta_{\mathcal{O}}|\Df)$ (times the OPE coefficients) at $h_0$. The operation of taking a residue commutes with the infinite sum because it can be expressed as a contour integral of a sum which converges uniformly along the contour. The individual term $\mathcal{I}(h;\Delta_{\mathcal{O}}|\Df)$ contributes two kinds of poles. First, this is the simple pole at $h=\Delta_{\mathcal{O}}$ with the right residue to precisely reproduce the operator $\mathcal{O}$ in the s-channel. Second, there are the double poles at double-trace dimensions $\Delta_n = 2\Df+2n+1$. Consistency with the s-channel OPE requires that $I_h$ has no singularity at these locations. This is only possible if the coefficient of both the single and double pole at $\Delta_n$ cancels out for all $n=0,1,\ldots$ when summed over all $\mathcal{O}$ in the OPE.

In order to write down the resulting sum rules on the OPE data, recall the OPE decomposition of the fermionic Polyakov block $P_{\Delta}(z)$
\be
P_{\Delta}(z) = G_{\Delta}(z) - \sum\limits_{n=0}^{\infty}\left[\alpha_n(\Delta)G_{\Delta_n}(z)+\beta_n(\Delta)\partial G_{\Delta_n}(z)\right]\,,
\label{eq:PBF}
\ee
where $\alpha_n(\Delta)$ and $\beta_n(\Delta)$ are computed from the inversion formula in \eqref{eq:abF}. The equations expressing the cancellation of the double-trace poles in $I_h$ read
\ba
&\sum\limits_{\mathcal{O}\in\phi\times\phi}\!\!(c_{\phi\phi\mathcal{O}})^2\,\alpha_n(\Delta_{\mathcal{O}}) = 0\quad\textrm{for }n=0,1,\ldots\\
&\sum\limits_{\mathcal{O}\in\phi\times\phi}\!\!(c_{\phi\phi\mathcal{O}})^2\,\beta_n(\Delta_{\mathcal{O}}) = 0\quad\textrm{for }n=0,1,\ldots\,.
\label{eq:eqsPolyakov}
\ea
What we have derived is that these equations must hold for the OPE data of every \emph{unitary} and \emph{crossing-symmetric} four-point function $\mathcal{G}(z)$. In the language of the Polyakov bootstrap in Mellin space, the first and second line in \eqref{eq:eqsPolyakov} express respectively the cancellation of the coefficient of the simple and double pole of the Mellin amplitude at the double-trace dimensions.

The equations hold even if there is a physical operator present in the OPE which sits precisely at a double-trace scaling dimension. The reason is that in that case the operator is still reproduced by the first term in \eqref{eq:PBF} and the square bracket gives a spurious contribution. For the same reason, the sums in \eqref{eq:eqsPolyakov} should include \emph{all} primary operators present in the OPE, including those precisely at double-trace dimensions.

One may wonder how fast the sums in \eqref{eq:eqsPolyakov} converge to zero. The answer is the same as for the rate of convergence of \eqref{eq:IFOPE} for a generic value of $h$. Namely, if we truncate the sums to operators with $\Delta_{\mathcal{O}}\leq\Delta_{\textrm{max}}$, the sums are $O(\Delta^{-2}_{\textrm{max}})$. We stress that for \eqref{eq:eqsPolyakov} to hold it is crucial that $\mathcal{G}(z)$ is bounded in the Regge limit. This is of course automatic if $\mathcal{G}(z)$ has a standard OPE decomposition into conformal blocks with positive coefficients. In more general cases, such as in perturbative $AdS$ theory where the OPE contains also derivatives of conformal blocks, we find that \eqref{eq:eqsPolyakov} still hold as long as $\mathcal{G}(z)$ is Regge-bounded.

The functions $\alpha_n(\Delta)$ and $\beta_n(\Delta)$ are in general quite complicated. Nevertheless, they have a very simple behaviour when $\Delta$ approaches the double-trace dimensions. Recall from Section \ref{sec:witten} that $P_{\Delta}(z)$ is the unique crossing-symmetric and Regge-bounded function whose $\dDisc$ agrees with $\dDisc$ of the t-channel conformal block of dimension $\Delta$. The latter $\dDisc$ has a second-order zero at every double-trace dimension. Since the zero function is crossing-symmetric, Regge bounded and has a zero $\dDisc$, also $P_{\Delta}(z)$ should have a second-order zero at all double-trace dimensions\footnote{$\dDisc$ of a t-channel conformal block has double zeros also at $\Delta=\Delta_n$ for $n$ negative. Our argument only applies for non-negative $n$ since $P_{\Delta}(z)$ is not normalizable for $\Delta<2\Df-1/2$.}
\be
P_{\Delta_n}(z) = 0\,,\quad \partial_{\Delta}P_{\Delta_n}(z) = 0\quad\textrm{for }n=0,1,\ldots\,.
\ee
Looking back at \eqref{eq:PBF}, these equations imply the following properties of the coefficients $\alpha_n(\Delta)$ and $\beta_n(\Delta)$ on the double-trace dimensions
\ba
&\alpha_n(\Delta_m) = \delta_{nm}\,,\qquad \partial_{\Delta}\alpha_n(\Delta_m) = 0\\
&\beta_n(\Delta_m) = 0\,,\qquad\quad\, \partial_{\Delta}\beta_n(\Delta_m) = \delta_{nm}\,,
\ea
where $n,m\in\mathbb{N}_{\geq0}$. In other words, $\beta_n(\Delta)$ has double zeros as a function of $\Delta$ on all the double traces except at $\Delta=\Delta_n$, where it has a simple zero with unit derivative. Similarly, $\alpha_n(\Delta)$ has double zeros on all the double traces except at $\Delta=\Delta_n$, where it equals one and has vanishing derivative.

As a final remark, note that we have not actually proven that the four-point function $\mathcal{G}(z)$ can be written as a sum of the Polyakov blocks since all our arguments took place at the level of the coefficient function. Indeed, in order to show that \eqref{eq:PBF} and \eqref{eq:eqsPolyakov} imply
\be
\mathcal{G}(z) = \!\!\sum\limits_{\mathcal{O}\in\phi\times\phi}\!\!(c_{\phi\phi\mathcal{O}})^2 P_{\Delta_{\mathcal{O}}}(z)\,,
\ee
we need to commute the infinite sums over $n$ and $\mathcal{O}$ in
\be
\sum\limits_{\mathcal{O}\in\phi\times\phi}\,
\sum\limits_{n=0}^{\infty}
(c_{\phi\phi\mathcal{O}})^2
\left[\alpha_n(\Delta)G_{\Delta_n}(z)+\beta_n(\Delta)\partial G_{\Delta_n}(z)\right]\stackrel{?}{=}0.
\ee
The argument that this is indeed possible is rather technical and is explained in Section 6 of the companion work \cite{Mazac:2018ycv} to which we refer the reader for details.

\subsection{Relationship to the analytic extremal functionals}\label{ssec:funtionals}
In the usual approach to the conformal bootstrap, one starts from the OPE \eqref{eq:gCalOPE} and derives constraints on the spectrum and OPE coefficients by imposing crossing
\be
\!\!\sum\limits_{\mathcal{O}\in\phi\times\phi}\!(c_{\phi\phi\mathcal{O}})^2F_{\Delta_{\mathcal{O}}}(z) = 0\,,
\label{eq:crossingF}
\ee
where
\be
F_{\Delta}(z)\equiv z^{-2\Df} G_{\Delta}(z) - (1-z)^{-2\Df} G_{\Delta}(1-z)\,.
\label{eq:fVector}
\ee
Sum rules satisfied by the OPE coefficients can be derived by applying linear functionals $\omega$ to \eqref{eq:crossingF}
\be
\!\!\sum\limits_{\mathcal{O}\in\phi\times\phi}\!(c_{\phi\phi\mathcal{O}})^2\omega[F_{\Delta_{\mathcal{O}}}] = 0\,.
\ee
$\omega$ must take finite values on all $F_{\Delta}(z)$ for $\Delta\geq0$, and must be swappable with the infinite sum over operators in \eqref{eq:crossingF}, see \cite{Rychkov:2017tpc,Mazac:2018mdx} for a detailed discussion of the swapping condition. The Polyakov bootstrap equations \eqref{eq:eqsPolyakov} also take the form of a sum rule on $(c_{\phi\phi\mathcal{O}})^2$ weighted by a nontrivial function of $\Delta_{\mathcal{O}}$. Therefore, we can derive \eqref{eq:eqsPolyakov} from the standard crossing equation provided we can construct linear functionals $\alpha^{\textrm{F}}_n$ and $\beta^{\textrm{F}}_n$ for $n\in\mathbb{N}_{\geq0}$ such that
\ba
\alpha^{\textrm{F}}_n\left[F_{\Delta}\right] &= \alpha^{\textrm{F}}_n(\Delta)\\
\beta^{\textrm{F}}_n\left[F_{\Delta}\right] &= \beta^{\textrm{F}}_n(\Delta)
\ea
for $\Delta\geq 0$. By a slight abuse of notation, $\alpha^{\textrm{F}}_n$ and $\beta^{\textrm{F}}_n$ on the LHS stand for the linear functional, whereas the same symbols on the RHS stand for the functions entering the OPE decomposition of the fermionic Polyakov blocks.

It turns out that linear functionals $\alpha^{\textrm{F}}_n$ and $\beta^{\textrm{F}}_n$ indeed exist. $\beta^{\textrm{F}}_0$ is the extremal functional for the maximization of the gap in $sl(2,\mathbb{R})$-invariant solutions to crossing, constructed in \cite{Mazac:2016qev,Mazac:2018mdx}. The remaining functionals, as well as their counterparts for the bosonic case were constructed in \cite{Mazac:2018ycv}. Below, we will quickly review the construction and show how it can be understood using our inversion formula.

The linear functionals of \cite{Mazac:2016qev,Mazac:2018mdx,Mazac:2018ycv} are each determined by a weight-function $f(z)$. Their action on a general function $\mathcal{F}(z)$ looks as follows
\be
\omega_f[\mathcal{F}] = \frac{1}{2}\!\!\!\int\limits_{\frac{1}{2}}^{\frac{1}{2}+i\infty}\!\!\!\!\!dz f(z)\mathcal{F}(z) \pm \!\int\limits_{\frac{1}{2}}^{1}\!\!dz\,(1-z)^{2\Df-2}f\!\left(\mbox{$\frac{z}{z-1}$}\right)\mathcal{F}(z)\,,
\ee
where here and in following the upper, lower sign applies in the bosonic, fermionic cases respectively. $f(z)$ must be holomorphic in the complex plane away from a branch cut at $z\in[0,1]$. $\omega_f$ defines a consistent functional which can be swapped with the OPE sum if and only if
\be
f(z) = O(z^{-1-\epsilon})
\label{eq:fAsym}
\ee
as $z\rightarrow\infty$ for some $\epsilon>0$. Furthermore, for the construction to work, $f(z)$ should be symmetric under crossing
\be
f(z) = f(1-z)
\label{eq:fSym}
\ee
and should satisfy the following functional equation
\be
z^{2\Df-2}f\!\left(\mbox{$\frac{1}{z}$}\right) + (1-z)^{2\Df-2}f\!\left(\mbox{$\frac{1}{1-z}$}\right) \pm \frac{f(z+i\epsilon)+f(z-i\epsilon)}{2} = 0
\label{eq:f3Term}
\ee
for $z\in(0,1)$. Under these assumptions, the functional action on the bootstrap vectors \eqref{eq:fVector} can be simplified as follows
\be
\omega[F_{\Delta}]=
\pm\!\!\int\limits_{0}^{1}\!\!dz z^{-2}f\!\left(\mbox{$\frac{1}{z}$}\right)\dDisc_{\textrm{B,F}}\!\left[G^{(t)}_{\Delta}(z)\right]\,,
\ee
or more explicitly
\ba
\omega_f[F_{\Delta}] &=
+2\sin^2\left[\frac{\pi}{2}(\Delta-2\Df)\right]
\!\!\int\limits_{0}^{1}\!\!dz z^{-2}f\!\left(\mbox{$\frac{1}{z}$}\right)\left(\mbox{$\frac{z}{1-z}$}\right)^{2\Df}\!\!G_\Delta(1-z)\\
\omega_f[F_{\Delta}] &=
-2\cos^2\left[\frac{\pi}{2}(\Delta-2\Df)\right]
\!\!\int\limits_{0}^{1}\!\!dz z^{-2}f\!\left(\mbox{$\frac{1}{z}$}\right)\left(\mbox{$\frac{z}{1-z}$}\right)^{2\Df}\!\!G_\Delta(1-z)\,,
\ea
where the first, second line corresponds to the bosonic, fermionic case respectively. Let us compare these formulas with the expressions for the OPE coefficients of bosonic and fermionic Polyakov blocks \eqref{eq:abB} and \eqref{eq:abF}. Clearly, they take the same form, with $f(z)$ related to the residues of the Lorentzian inversion kernel at the appropriate double trace dimension. Specifically, to construct the functionals $\alpha^{\textrm{B}}_n$, $\beta^{\textrm{B}}_n$ computing the OPE expansion of the bosonic Polyakov blocks, we should take (for $\Df\notin\mathbb{N}$)
\ba
\alpha^{\textrm{B}}_n:\quad f(z) &= \widehat{H}^{\textrm{B}}_{n,1}\!\left(\mbox{$\frac{1}{z}$}\right)\\
\beta^{\textrm{B}}_n:\quad f(z) &= \widehat{H}^{\textrm{B}}_{n,2}\!\left(\mbox{$\frac{1}{z}$}\right)\,,
\label{eq:fFromHB1}
\ea
where $\widehat{H}^{\textrm{B}}_{n,1}(z)$ and $\widehat{H}^{\textrm{B}}_{n,2}(z)$ are defined from the bosonic inversion kernel in equation \eqref{eq:hHatBDef}. When $\Df\in\mathbb{N}$, the above formula needs to be corrected by the contribution of the discrete series
\ba
\alpha^{\textrm{B}}_n:\quad f(z) &= \widehat{H}^{\textrm{B}}_{n,1}\!\left(\mbox{$\frac{1}{z}$}\right)
- \frac{2\Gamma(\Delta^{\textrm{B}}_n)^4}{\pi^2\Gamma(2\Delta^{\textrm{B}}_n)\Gamma(2\Delta^{\textrm{B}}_n-1)}G_{\Delta^{\textrm{B}}_n}\!\left(\mbox{$\frac{1}{z}$}\right)\\
\beta^{\textrm{B}}_n:\quad f(z) &= \widehat{H}^{\textrm{B}}_{n,2}\!\left(\mbox{$\frac{1}{z}$}\right)\,.
\label{eq:fFromHB2}
\ea
Similarly, in the fermionic case we find for $\Df\notin\mathbb{N}-\frac{1}{2}$
\ba
\alpha^{\textrm{F}}_n:\quad f(z) &= -\widehat{H}^{\textrm{F}}_{n,1}\!\left(\mbox{$\frac{1}{z}$}\right)\\
\beta^{\textrm{F}}_n:\quad f(z) &= -\widehat{H}^{\textrm{F}}_{n,2}\!\left(\mbox{$\frac{1}{z}$}\right)\,,
\ea
and for $\Df\in\mathbb{N}-\frac{1}{2}$
\ba
\alpha^{\textrm{F}}_n:\quad f(z) &= -\widehat{H}^{\textrm{F}}_{n,1}\!\left(\mbox{$\frac{1}{z}$}\right)
+ \frac{2\Gamma(\Delta^{\textrm{F}}_n)^4}{\pi^2\Gamma(2\Delta^{\textrm{F}}_n)\Gamma(2\Delta^{\textrm{F}}_n-1)}G_{\Delta^{\textrm{F}}_n}\!\left(\mbox{$\frac{1}{z}$}\right)\\
\beta^{\textrm{F}}_n:\quad f(z) &= -\widehat{H}^{\textrm{F}}_{n,2}\!\left(\mbox{$\frac{1}{z}$}\right)\,.
\ea
with $\widehat{H}^{\textrm{F}}_{n,1}(z)$ and $\widehat{H}^{\textrm{F}}_{n,2}(z)$ defined in \eqref{eq:hHatFDef}.

The connection to the Lorentzian inversion formula gives an alternative explanation why $f(z)$ satisfies the constraints \eqref{eq:fSym}, \eqref{eq:f3Term}. Indeed \eqref{eq:fSym} follows immediately from the symmetry of the inversion kernels under $z\mapsto\mbox{$\frac{z}{z-1}$}$, i.e. equation \eqref{eq:HSym}. The three-term identity \eqref{eq:f3Term} follows from the functional equation \eqref{eq:feIH} satisfied by $H^{\textrm{B,F}}_{\Delta}(z)$ in order for the Euclidean and Lorentzian inversion formulas to be compatible. Now, $H^{\textrm{B,F}}_{\Delta}(z)$ satisfies the identity with a nonzero RHS. Recall that $f\!\left(\mbox{$\frac{1}{z}$}\right)$ is proportional to the residues of $H^{\textrm{B,F}}_{\Delta}(z)$ at $\Delta=\Delta^{\textrm{B,F}}_n$. Since the RHS of \eqref{eq:feIH} is finite at these values of $\Delta$, we can derive the identity for $f(z)$ from the identity for $H^{\textrm{B,F}}_{\Delta}(z)$ by taking the residues of the latter at the double-trace dimensions.

Since the fermionic inversion kernel satisfies $H^{\textrm{F}}_{\Delta}(z) = O(z^{2})$ as $z\rightarrow 0$, the resulting $f(z)$ will have the asymptotic behaviour \eqref{eq:fAsym}, necessary for the functional to satisfy the swapping condition. On the other hand, the bosonic inversion kernels $H^{\textrm{B}}_{\Delta}(z)=O(z^0)$ as $z\rightarrow 0$, which is in general not enough for the swapping condition to be satisfied. This is equivalent to saying that the bosonic inversion formula with kernel $H^{\textrm{B}}_{\Delta}$ only holds for super-bounded four-point functions. In the next subsection, we will amend this shortcoming.

We have explained how the Polyakov bootstrap equations are a consequence of the standard crossing equation. One may wonder whether one can also go in the opposite direction and show that whenever a set of putative OPE data $\{\Delta_{\mathcal{O}},(c_{\phi\phi\mathcal{O}})^2\}_{\mathcal{O}\in\phi\times\phi}$ satisfies \eqref{eq:eqsPolyakov} for all $n\in\mathbb{N}_{\geq0}$, then it automatically leads to a crossing-symmetric four-point function. The answer is affirmative, as proven in Section 6 of \cite{Mazac:2018ycv}.

\subsection{Improved bosonic inversion formula}\label{ssec:reggeRelax}
We will now complete the picture by deriving a Lorentzian inversion formula for the bosonic case which applies to general Regge-bounded four-point functions, and not just the super-bounded ones.

Recall that the Regge behaviour of the four-point function is detected by the $z\rightarrow0$ limit of the Lorentzian inversion kernel. In particular, to have an inversion formula for the bosonic case which applies to all Regge-bounded correlators, we must make sure $H_{\Delta}(z) = O(z^2)$ as $z\rightarrow 0$, while $H^{\textrm{B}}_\Delta(z)$ of Section \ref{sec:HFormulas} only satisfies the weaker constraint $H^{\textrm{B}}_\Delta(z) = O(z^0)$. We will construct $H_{\Delta}(z)$ by starting with $H^{\textrm{B}}_\Delta(z)$ and subtracting from it a correcting function $H^{\textrm{cor}}_{\Delta}(z)$ such that $H_{\Delta}(z)=H^{\textrm{B}}_\Delta(z) - H^{\textrm{cor}}_{\Delta}(z) = O(z^2)$ as $z\rightarrow 0$. The resulting kernel $H_{\Delta}(z)$ still needs to satisfy the symmetry condition \eqref{eq:HSym} and the functional equation \eqref{eq:feIH}. This means that $H^{\textrm{cor}}_{\Delta}(z)$ must satisfy the same symmetry condition, and also the functional equation with zero RHS. We have seen examples of functions satisfying these constraints in the previous subsection. These are the coefficients of the double or simple pole of $H^{\textrm{B}}_{\Delta}(z)$ at double-trace $\Delta$, which we denoted $\widehat{H}^{\textrm{B}}_{n,2}(z)$ and $\widehat{H}^{\textrm{B}}_{n,1}(z)$ and defined in \eqref{eq:hHatBDef}. We will work with $\widehat{H}^{\textrm{B}}_{0,2}(z)$, i.e. the coefficient of the double pole at $\Delta = 2\Df$, for simplicity. Its expansion for $z\rightarrow0$ starts with
\be
\widehat{H}^{\textrm{B}}_{0,2}(z) = 
\frac{4 \Gamma (\Df)^2 \Gamma \left(2\Df-\frac{1}{2}\right)}{\pi ^{3/2}\Gamma (2 \Df)\Gamma \left(\Df-\frac{1}{2}\right)^2}+O(z^2)\,.
\ee
Therefore we should set
\be
H_{\Delta}(z) \equiv H^{\textrm{B}}_{\Delta}(z) -
\frac{\pi ^2 2^{2(\Df-1)} \Gamma \left(\Df+\frac{1}{2}\right)}{\Gamma (\Df)^3 \Gamma \left(2\Df-\frac{1}{2}\right)}
\frac{\Gamma \left(\Df-\frac{\Delta}{2}\right)^2 \Gamma \left(\Df-\frac{1-\Delta}{2}\right)^2}{\Gamma \left(1-\frac{\Delta}{2}\right)^2 \Gamma \left(1-\frac{1-\Delta}{2}\right)^2}\frac{2\pi}{\sin (\pi  \Delta)}\widehat{H}^{\textrm{B}}_{0,2}(z)\,,
\label{eq:hbCorrected}
\ee
where we used the series expansion of $H_{\Delta}^{\textrm{B}}(z)$ given in \eqref{eq:HBGeneral} to guarantee $H_{\Delta}(z) = O(z^2)$.

While the subtraction of $\widehat{H}^{\textrm{B}}_{0,2}(z)$ cured the behaviour of the inversion kernel as $z\rightarrow 0$, it introduced a new singularity at $z\rightarrow 1$. Indeed, it follows from \eqref{eq:z1Power} that
\be
\widehat{H}^{\textrm{B}}_{0,2}(z) \stackrel{z\rightarrow 1}{\sim}\frac{2}{\pi^2}\frac{1}{1-z}\,.
\ee
This singularity implies that the Lorentzian inversion integral which uses $H_{\Delta}(z)$ as the kernel would diverge if there were primary operators $\mathcal{O}$ with $\Delta_{\mathcal{O}}<2\Df$ contributing to $\dDisc$ via the t-channel OPE. We can cure this problem as before by subtracting from $\mathcal{G}(z)$ the finitely many bosonic Polyakov blocks of operators with $\Delta_{\mathcal{O}}\in[0,2\Df)$ to get the regularized correlator $\mathcal{G}^{\textrm{reg}}(z)$. However, this introduces additional double-trace contributions in the OPE. Although these do not contribute to the double discontinuity, they can contribute in the intermediate steps relating the Euclidean and Lorentzian inversion formula. Keeping track of the extra terms, we find that the correct answer for the coefficient function is
\ba
I^{\textrm{reg}}_{\Delta} =&
2\int\limits_{0}^{1}\!\! dz z^{-2}H_{\Delta}(z)\dDisc_{\textrm{B}}\!\left[\mathcal{G}^{\textrm{reg}}(z)\right] +\\
&+\lim_{\epsilon\rightarrow 0}\int\limits_{C^{+}_\epsilon}\!\!
dz z^{-2}H_{\Delta}(z)\mathcal{G}^{\textrm{reg}}(z)+
\lim_{\epsilon\rightarrow 0}
\int\limits_{C^{-}_\epsilon}\!\!
dz z^{-2}H_{\Delta}(z)\mathcal{G}^{\textrm{reg}}(z)\,,
\label{eq:libcore}
\ea
where $C_{\epsilon}^{\pm}$ are semicircular contours of radius $\epsilon$ going from $z=1+\epsilon$ to $z=1-\epsilon$ above and below the real axis respectively. $\mathcal{G}^{\textrm{reg}}(z)$ in the last two terms is the analytic continuation of $\mathcal{G}^{\textrm{reg}}(z)$ from the region $z\in(1,\infty)$.

As a special case, we can apply this formula to the scalar contact diagram in $AdS_2$ with no derivatives to find its coefficient function. The diagram is crossing-symmetric and bounded in the Regge limit. Therefore the formula applies to it and the first term vanishes since the double discontinuity of the contact diagram vanishes. Therefore, the entire coefficient function comes from the infinitesimal contour integrals. These integrals in turn only receive a contribution from the simple pole of $H_{\Delta}(z)$ at $z=1$, which is entirely due to the second term in \eqref{eq:hbCorrected}. The integrals are only sensitive to the anomalous dimension of the leading double-trace $\gamma_0$ in the contact diagram, which therefore normalizes the whole diagram. The result is
\be
I^{\textrm{contact}}_{\Delta} =
-\frac{2 \pi ^{5/2} \Gamma (2\Df)}{\Gamma (\Df)^4 \Gamma \left(2\Df-\frac{1}{2}\right)}
\frac{\Gamma \left(\Df-\frac{\Delta}{2}\right)^2 \Gamma \left(\Df-\frac{1-\Delta}{2}\right)^2}{\Gamma \left(1-\frac{\Delta}{2}\right)^2 \Gamma \left(1-\frac{1-\Delta}{2}\right)^2}\frac{2\pi}{\sin (\pi  \Delta)}\times\gamma_0\,,
\ee
which agrees with the correct answer.

Let us now apply the inversion formula \eqref{eq:libcore} to a general physical four-point function and expand it in the t-channel. It is not hard to check that analogously to the fermionic result \eqref{eq:IFOPE}, we get
\be
I_h = \!\!\sum\limits_{\mathcal{O}\in\phi\times\phi}\!
(c_{\phi\phi\mathcal{O}})^2\,\mathcal{I}(h;\Delta_{\mathcal{O}}|\Df)\,,
\ee
where
\be
\mathcal{I}(h;\Delta_{\mathcal{O}}|\Df) = \mathcal{I}_{\textrm{B}}(h;\Delta_{\mathcal{O}}|\Df) - \lambda(\Delta_{\mathcal{O}};\Df) I^{\textrm{contact}}_{h}\,,
\ee
and $\lambda(\Delta_{\mathcal{O}};\Df)$ is chosen such that $\mathcal{I}(h;\Delta_{\mathcal{O}}|\Df)$ exhibits no double pole at $h=2\Df$. The reason is that the second term in the improved inversion kernel \eqref{eq:hbCorrected} makes the double pole of $H_{\Delta}^{\textrm{B}}(z)$ at $\Delta=2\Df$ into a simple pole.

This means that all the main conclusions derived in the fermionic case in \ref{ssec:polyakovArgument} and \ref{ssec:sumrules} are still valid in the bosonic case provided we use the following definition of bosonic Polyakov blocks: they are the crossing-symmetric sum of exchange Witten diagrams in the s-, t- and u-channel plus the non-derivative scalar contact diagram whose coefficient is chosen to precisely cancel the term $\partial_\Delta G_{2\Df}(z)$ in the conformal block expansion of the sum of exchange diagrams.

This definition is not canonical as we could have chosen a different $\widehat{H}^{\textrm{B}}_{n,2}(z)$ or $\widehat{H}^{\textrm{B}}_{n,1}(z)$ to improve the $z\rightarrow 0$ behaviour of the inversion kernel. This would lead to a definition of the bosonic Polyakov blocks where the contact diagram is chosen to cancel any single double-trace conformal block or its $\Delta$-derivative. Any such definition would give a set of sum rules which are entirely equivalent to the standard crossing equation for Regge-bounded four-point function \cite{Mazac:2018ycv}.

\section{Discussion and open questions}\label{sec:discussion}
The main result of this note is a Lorentzian inversion formula for the decomposition of conformal four-point functions into the principal series of the 1D conformal group. The formula expresses the coefficient function $I_{\Delta}$ as an integral of the double discontinuity of the correlator times an inversion kernel $H_{\Delta}(z)$. It is analogous to the Lorentzian inversion formula of Caron-Huot \cite{Caron-Huot2017b}, which applies to the principal series of the conformal group in more than one dimension.

The formula of this note was derived by a contour deformation from the Euclidean inversion formula. The contour-deformation argument is valid only for Regge-bounded four-point functions of \emph{identical} bosons or fermions. The argument does not yield an explicit formula for $H_{\Delta}(z)$ but only a functional equation that $H_{\Delta}(z)$ must satisfy in order for the Lorentzian and Euclidean formulas to agree. A closed-form solution for $H_{\Delta}(z)$ was presented in a number of cases. The functional equation and therefore also $H_{\Delta}(z)$ depend on the external dimension $\Df$ in a rather nontrivial way. This is a price we need to pay for having a formula which manifests crossing symmetry, since the latter is a non-trivial property whose form also depends on $\Df$.

The inversion formula manifests crossing symmetry in the following sense. We can apply it to a crossing-symmetric four-point function, and expand the latter into t-channel conformal blocks inside the formula. When applied to an individual t-channel conformal block of dimension $\Delta_{\mathcal{O}}$, the formula returns the coefficient function of the crossing-symmetric sum of exchange Witten diagrams in $AdS_2$ in the s-, t- and u-channel. Crucially, this includes the s-channel exchange diagram which itself includes the s-channel conformal block of dimension $\Delta_{\mathcal{O}}$ with unit coefficient. In other words, by inserting a primary operator into the t-channel, we get it back in the s-channel dressed by double-trace contributions which make the result crossing-symmetric.

It follows that inserting the t-channel conformal block expansion into the inversion formula gives the coefficient function $I_{\Delta}$ as an infinite sum over the coefficient functions of crossing-symmetric exchange Witten diagrams. The sum converges for any complex $\Delta$ away from the poles of the individual summands, and in particular everywhere away from the real axis. Since the summands are meromorphic functions of $\Delta$ and the convergence is uniform, this proves that $I_{\Delta}$ is meromorphic in the entire complex plane.

Let us contrast the above properties with those of the standard $D>1$ inversion formula of Caron-Huot. The $D>1$ formula neither requires nor manifests crossing symmetry. Indeed, it applies to arbitrary four-point functions of operators which may or may not be identical. The inversion kernel is an s-channel conformal block with Weyl-reflected quantum numbers -- considerably simpler than $H_{\Delta}(z)$ which are needed for the crossing-symmetric formula of this note.

Just like before, Caron-Huot's formula allows us to compute the s-channel coefficient function $I_{\Delta,J}$ in terms of the OPE in the crossed channels. To make the comparison clear, let us apply Caron-Huot's formula to a four-point function of identical operators. If we insert a single conformal block in the t- and u-channels, we get back the coefficient function of the sum of exchange diagrams in the t- and u-channel \cite{Liu:2018jhs} (possibly supplemented by contact diagrams). This is not a fully crossing-symmetric object. In particular, the s-channel exchange diagram is absent, unlike when using the formula of this note. This means that inserting an individual primary operator in the crossed channels does not produce the same operator in the s-channel, but instead only an infinite tower of double-trace contributions. The double-trace poles appear as a result of performing the $z$-integral in Caron-Huot's formula. On the other hand, we saw that in our formula the analogous poles are present already in $H_{\Delta}(z)$, i.e. even before performing the $z$-integral.

How then does the $D>1$ formula reproduce the poles of $I_{\Delta,J}$ at the correct locations of s-channel primary operators? Such poles (and therefore exact crossing symmetry) can only arise from summing over infinitely many operators in the crossed-channel OPEs. Indeed, the crossed-channel sum representation of $I_{\Delta,J}$ will only converge in a finite strip around the principal series where $\mathrm{Re}(\Delta)$ is smaller than the dimension of the first s-channel primary of a given spin. Again, this is very different from what we found for the inversion formula of this note, where the crossed-channel sum for $I_{\Delta}$ converged everywhere away from the expected poles.

In spite of the described differences between the $D=1$ formula of this note and the $D>1$ formula of Caron-Huot, the former is equally useful for implementing analytic conformal bootstrap. The conformal bootstrap constraints arise in our context in the following way. Recall that our inversion formula leads to an expansion of the four-point function similar to the standard s-channel conformal block expansion, except each conformal block gets replaced by the crossing-symmetric sum of exchange diagrams. Consistency with the standard OPE then requires that the contributions at non-interacting double-trace dimensions must drop out when summed over all physical operators in the OPE. Therefore, we get an infinite set of sum rules satisfied by the OPE data, labelled by the double-trace operators. This idea was first introduced by Polyakov in \cite{Polyakov:1974gs} and developed in recent works \cite{Sen:2015doa,Gopakumar2017,Gopakumar2017a,Dey:2017fab,Dey:2016mcs,Gopakumar:2018xqi}.

Another recent work \cite{Mazac:2018ycv}, closely related to the present note, derived these sum rules by applying distinguished linear functionals to the standard crossing equation, showing in particular that these sum rules are a completely equivalent reformulation of the standard crossing equation. In the main text, we gave an explanation of how the relevant functionals arise from the Lorentzian inversion formula. These functionals are examples of optimal functionals of the numerical bootstrap. Therefore the 1D inversion formula clarifies how the numerical, analytical and Polyakov's approach to the conformal bootstrap are connected.

An important application of Caron-Huot's formula has been the perturbative expansion of the CFT data around mean field theory \cite{Caron-Huot2017b,Alday:2017zzv,Aharony:2018npf}. Equivalently, this can be viewed as the computation of loop-level Witten diagrams in $AdS$ from crossing symmetry \cite{Aharony:2016dwx,Alday:2017xua,Alday:2017vkk,Alday:2018pdi}. The effectiveness of the inversion formula in this context stems from the fact that the double discontinuity at one loop is fixed entirely in terms of tree-level OPE data. The $D=1$ inversion formula can be used in exactly the same way to compute loop-level diagrams in $AdS_2$. The computation was carried out up to two loops using the equivalent language of bootstrap functionals in Section 7 of \cite{Mazac:2018ycv}.

This work leaves a number of open questions and future directions to be explored. At the practical level, it would be desirable to find a closed formula for the inversion kernels $H_{\Delta}(z)$ valid for general $\Df$, starting either from the functional equation of Section \ref{ssec:LorConstraint} or otherwise. One could then apply the inversion formula to a single t-channel block to find the OPE decomposition of crossing-symmetrized Witten diagrams for general $\Df$ and $\Delta_{\mathcal{O}}$. The latter task may perhaps be achieved independently using the techniques of \cite{Sleight:2018ryu} or \cite{Zhou:2018sfz}.

We saw our inversion formula behaves differently from Caron-Huot's formula in some important aspects. Does it mean that there is an alternative Lorentzian inversion formula in $D>1$ which is more closely analogous to the $D=1$ formula of this note? Such formula would apply to four-point functions of identical primaries and should have the property that the inverse of a single crossed-channel block is a crossing-symmetric object which contains the corresponding single-trace pole in the s-channel. There is a reason to doubt the existence of such formula in $D>1$, having to do Regge-boundedness. If the formula existed, the inverse of a crossed-channel conformal block would presumably be a Regge-bounded and crossing-symmetric combination of exchange diagrams and contact diagrams. However, such combination does not exist if the exchanged spin is greater than one. The reason is that if it existed, the standard Lorentzian inversion formula would apply to it and thanks to Regge-boundedness would give the correct s-channel OPE data down to (and including) spin two. Therefore, the s-channel data would have to include only double-trace contributions down to spin two, giving a contradiction with crossing symmetry, which requires the single-trace pole in the s-channel. 

Nevertheless, one can hope that progress can be made by relaxing some of the conditions the modified inversion formula should satisfy. This line of thought is very interesting since it could lead to a better analytic understanding of the optimal bounds of the $D>1$ numerical bootstrap, as well as the $D>1$ Polyakov-Mellin bootstrap in the fashion of \cite{Mazac:2018ycv} or the present work.

It would be interesting to generalize the presented formula to the case of non-identical external operators. To implement full crossing symmetry, the input of such formula should consist of all four-point functions that one can construct with a given set of external operators. Its output would then be all coefficient functions of this set of correlators. Furthermore, one should look for a more conceptual derivation of the 1D formula along the lines of \cite{Kravchuk:2018htv}.

We gave a proof that $I_{\Delta}$ of a four-point function in a unitary 1D theory is a meromorphic function of $\Delta$. It is expected that $I_{\Delta,J}$ in $D>1$ satisfies the same property, but it would be rewarding to find a proof of this statement, especially for general complex $J$.

It should be straightforward to modify the results of this note to 1D theories with superconformal symmetry, such as the half-BPS Wilson loop in $\mathcal{N}=4$ SYM \cite{Giombi:2018qox,Giombi:2017cqn,Giombi:2018hsx,Liendo:2018ukf}. In particular, one could use it to perform the strong-coupling expansion of the CFT data using crossing symmetry.

Finally, there exists an intriguing connection between the present work and the recent solution of the sphere-packing problem in 8 and 24 dimensions \cite{2016arXiv160304246V,2016arXiv160306518C}. Indeed, the method of \cite{2016arXiv160304246V,2016arXiv160306518C} is essentially identical to the construction of analytic extremal functional given in \cite{Mazac:2016qev} and \cite{Mazac:2018mdx}, modified to the context of modular bootstrap in the presence of an abelian current algebra. This connection will be explored in more detail in \cite{ModularSphere}.

\subsection*{Acknowledgements}
The author would like to thank Simon Caron-Huot, Rajesh Gopakumar, Zohar Komargodski, Shota Komatsu, Petr Kravchuk, Miguel Paulos, Leonardo Rastelli, Aninda Sinha and Xinan Zhou for useful discussions and/or comments on the draft. He is also grateful to the organizers of the May 2018 Bootstrap collaboration workshop on Analytic approaches to the bootstrap for creating a stimulating atmosphere during the initial stages of this work.

\small
\parskip=-10pt
\bibliography{Functionals}

\providecommand{\href}[2]{#2}\begingroup\raggedright\begin{thebibliography}{10}

\bibitem{Rattazzi:2008pe}
R.~Rattazzi, V.~S. Rychkov, E.~Tonni, and A.~Vichi, {\it {Bounding scalar
  operator dimensions in 4D CFT}},  {\em JHEP} {\bf 12} (2008) 031,
  [\href{http://arxiv.org/abs/0807.0004}{{\tt arXiv:0807.0004}}].

\bibitem{ElShowk:2012ht}
S.~El-Showk, M.~F. Paulos, D.~Poland, S.~Rychkov, D.~Simmons-Duffin, and
  A.~Vichi, {\it {Solving the 3D Ising Model with the Conformal Bootstrap}},
  {\em Phys. Rev.} {\bf D86} (2012) 025022,
  [\href{http://arxiv.org/abs/1203.6064}{{\tt arXiv:1203.6064}}].

\bibitem{Ferrara:1973yt}
S.~Ferrara, A.~F. Grillo, and R.~Gatto, {\it {Tensor representations of
  conformal algebra and conformally covariant operator product expansion}},
  {\em Annals Phys.} {\bf 76} (1973) 161--188.

\bibitem{Polyakov:1974gs}
A.~M. Polyakov, {\it {Nonhamiltonian approach to conformal quantum field
  theory}},  {\em Zh. Eksp. Teor. Fiz.} {\bf 66} (1974) 23--42. [Sov. Phys.
  JETP39,9(1974)].

\bibitem{Mack:1975jr}
G.~Mack, {\it {Duality in quantum field theory}},  {\em Nucl. Phys.} {\bf B118}
  (1977) 445--457.

\bibitem{Poland:2018epd}
D.~Poland, S.~Rychkov, and A.~Vichi, {\it {The Conformal Bootstrap: Theory,
  Numerical Techniques, and Applications}},
  \href{http://arxiv.org/abs/1805.04405}{{\tt arXiv:1805.04405}}.

\bibitem{Komargodski:2012ek}
Z.~Komargodski and A.~Zhiboedov, {\it {Convexity and Liberation at Large
  Spin}},  {\em JHEP} {\bf 11} (2013) 140,
  [\href{http://arxiv.org/abs/1212.4103}{{\tt arXiv:1212.4103}}].

\bibitem{Fitzpatrick:2012yx}
A.~L. Fitzpatrick, J.~Kaplan, D.~Poland, and D.~Simmons-Duffin, {\it {The
  Analytic Bootstrap and AdS Superhorizon Locality}},  {\em JHEP} {\bf 12}
  (2013) 004, [\href{http://arxiv.org/abs/1212.3616}{{\tt arXiv:1212.3616}}].

\bibitem{Alday:2015ewa}
L.~F. Alday and A.~Zhiboedov, {\it {An Algebraic Approach to the Analytic
  Bootstrap}},  {\em JHEP} {\bf 04} (2017) 157,
  [\href{http://arxiv.org/abs/1510.08091}{{\tt arXiv:1510.08091}}].

\bibitem{Alday:2016njk}
L.~F. Alday, {\it {Large Spin Perturbation Theory for Conformal Field
  Theories}},  {\em Phys. Rev. Lett.} {\bf 119} (2017), no.~11 111601,
  [\href{http://arxiv.org/abs/1611.01500}{{\tt arXiv:1611.01500}}].

\bibitem{Caron-Huot2017b}
S.~Caron-Huot, {\it {Analyticity in Spin in Conformal Theories}},  {\em JHEP}
  {\bf 09} (2017) 078, [\href{http://arxiv.org/abs/1703.00278}{{\tt
  arXiv:1703.00278}}].

\bibitem{Simmons-Duffin:2017nub}
D.~Simmons-Duffin, D.~Stanford, and E.~Witten, {\it {A spacetime derivation of
  the Lorentzian OPE inversion formula}},
  \href{http://arxiv.org/abs/1711.03816}{{\tt arXiv:1711.03816}}.

\bibitem{Kravchuk:2018htv}
P.~Kravchuk and D.~Simmons-Duffin, {\it {Light-ray operators in conformal field
  theory}},  \href{http://arxiv.org/abs/1805.00098}{{\tt arXiv:1805.00098}}.

\bibitem{Mukhametzhanov:2018zja}
B.~Mukhametzhanov and A.~Zhiboedov, {\it {Analytic Euclidean Bootstrap}},
  \href{http://arxiv.org/abs/1808.03212}{{\tt arXiv:1808.03212}}.

\bibitem{Alday:2017zzv}
L.~F. Alday, J.~Henriksson, and M.~van Loon, {\it {Taming the
  $\epsilon$-expansion with large spin perturbation theory}},  {\em JHEP} {\bf
  07} (2018) 131, [\href{http://arxiv.org/abs/1712.02314}{{\tt
  arXiv:1712.02314}}].

\bibitem{Aharony:2018npf}
O.~Aharony, L.~F. Alday, A.~Bissi, and R.~Yacoby, {\it {The Analytic Bootstrap
  for Large $N$ Chern-Simons Vector Models}},  {\em JHEP} {\bf 08} (2018) 166,
  [\href{http://arxiv.org/abs/1805.04377}{{\tt arXiv:1805.04377}}].

\bibitem{Aharony:2016dwx}
O.~Aharony, L.~F. Alday, A.~Bissi, and E.~Perlmutter, {\it {Loops in AdS from
  Conformal Field Theory}},  {\em JHEP} {\bf 07} (2017) 036,
  [\href{http://arxiv.org/abs/1612.03891}{{\tt arXiv:1612.03891}}].

\bibitem{Alday:2017xua}
L.~F. Alday and A.~Bissi, {\it {Loop Corrections to Supergravity on $AdS_5
  \times S^5$}},  {\em Phys. Rev. Lett.} {\bf 119} (2017), no.~17 171601,
  [\href{http://arxiv.org/abs/1706.02388}{{\tt arXiv:1706.02388}}].

\bibitem{Alday:2017vkk}
L.~F. Alday and S.~Caron-Huot, {\it {Gravitational S-matrix from CFT dispersion
  relations}},  \href{http://arxiv.org/abs/1711.02031}{{\tt arXiv:1711.02031}}.

\bibitem{Alday:2018pdi}
L.~F. Alday, A.~Bissi, and E.~Perlmutter, {\it {Genus-One String Amplitudes
  from Conformal Field Theory}},  \href{http://arxiv.org/abs/1809.10670}{{\tt
  arXiv:1809.10670}}.

\bibitem{Billo:2013jda}
M.~Bill\'{o}, M.~Caselle, D.~Gaiotto, F.~Gliozzi, M.~Meineri, and
  R.~Pellegrini, {\it {Line defects in the 3d Ising model}},  {\em JHEP} {\bf
  07} (2013) 055, [\href{http://arxiv.org/abs/1304.4110}{{\tt
  arXiv:1304.4110}}].

\bibitem{Gaiotto:2013nva}
D.~Gaiotto, D.~Mazac, and M.~F. Paulos, {\it {Bootstrapping the 3d Ising twist
  defect}},  {\em JHEP} {\bf 03} (2014) 100,
  [\href{http://arxiv.org/abs/1310.5078}{{\tt arXiv:1310.5078}}].

\bibitem{Giombi:2018qox}
S.~Giombi and S.~Komatsu, {\it {Exact Correlators on the Wilson Loop in
  $\mathcal{N}=4$ SYM: Localization, Defect CFT, and Integrability}},
  \href{http://arxiv.org/abs/1802.05201}{{\tt arXiv:1802.05201}}.

\bibitem{Giombi:2017cqn}
S.~Giombi, R.~Roiban, and A.~A. Tseytlin, {\it {Half-BPS Wilson loop and
  AdS$_2$/CFT$_1$}},  {\em Nucl. Phys.} {\bf B922} (2017) 499--527,
  [\href{http://arxiv.org/abs/1706.00756}{{\tt arXiv:1706.00756}}].

\bibitem{Giombi:2018hsx}
S.~Giombi and S.~Komatsu, {\it {More Exact Results in the Wilson Loop Defect
  CFT: Bulk-Defect OPE, Nonplanar Corrections and Quantum Spectral Curve}},
  \href{http://arxiv.org/abs/1811.02369}{{\tt arXiv:1811.02369}}.

\bibitem{Liendo:2018ukf}
P.~Liendo, C.~Meneghelli, and V.~Mitev, {\it {Bootstrapping the half-BPS line
  defect}},  {\em JHEP} {\bf 10} (2018) 077,
  [\href{http://arxiv.org/abs/1806.01862}{{\tt arXiv:1806.01862}}].

\bibitem{Sachdev:1992fk}
S.~Sachdev and J.~Ye, {\it {Gapless spin fluid ground state in a random,
  quantum Heisenberg magnet}},  {\em Phys. Rev. Lett.} {\bf 70} (1993) 3339,
  [\href{http://arxiv.org/abs/cond-mat/9212030}{{\tt cond-mat/9212030}}].

\bibitem{Maldacena:2016hyu}
J.~Maldacena and D.~Stanford, {\it {Remarks on the Sachdev-Ye-Kitaev model}},
  {\em Phys. Rev.} {\bf D94} (2016), no.~10 106002,
  [\href{http://arxiv.org/abs/1604.07818}{{\tt arXiv:1604.07818}}].

\bibitem{Caron-Huot:2018kta}
S.~Caron-Huot and A.-K. Trinh, {\it {All Tree-Level Correlators in
  AdS${}_5\times$S${}_5$ Supergravity: Hidden Ten-Dimensional Conformal
  Symmetry}},  \href{http://arxiv.org/abs/1809.09173}{{\tt arXiv:1809.09173}}.

\bibitem{Roberts:2014ifa}
D.~A. Roberts and D.~Stanford, {\it {Two-dimensional conformal field theory and
  the butterfly effect}},  {\em Phys. Rev. Lett.} {\bf 115} (2015), no.~13
  131603, [\href{http://arxiv.org/abs/1412.5123}{{\tt arXiv:1412.5123}}].

\bibitem{Maldacena:2015waa}
J.~Maldacena, S.~H. Shenker, and D.~Stanford, {\it {A bound on chaos}},  {\em
  JHEP} {\bf 08} (2016) 106, [\href{http://arxiv.org/abs/1503.01409}{{\tt
  arXiv:1503.01409}}].

\bibitem{Perlmutter:2016pkf}
E.~Perlmutter, {\it {Bounding the Space of Holographic CFTs with Chaos}},  {\em
  JHEP} {\bf 10} (2016) 069, [\href{http://arxiv.org/abs/1602.08272}{{\tt
  arXiv:1602.08272}}].

\bibitem{Sen:2015doa}
K.~Sen and A.~Sinha, {\it {On critical exponents without Feynman diagrams}},
  {\em J. Phys.} {\bf A49} (2016), no.~44 445401,
  [\href{http://arxiv.org/abs/1510.07770}{{\tt arXiv:1510.07770}}].

\bibitem{Gopakumar2017}
R.~Gopakumar, A.~Kaviraj, K.~Sen, and A.~Sinha, {\it {Conformal Bootstrap in
  Mellin Space}},  {\em Phys. Rev. Lett.} {\bf 118} (2017), no.~8 081601,
  [\href{http://arxiv.org/abs/1609.00572}{{\tt arXiv:1609.00572}}].

\bibitem{Gopakumar2017a}
R.~Gopakumar, A.~Kaviraj, K.~Sen, and A.~Sinha, {\it {A Mellin space approach
  to the conformal bootstrap}},  {\em JHEP} {\bf 05} (2017) 027,
  [\href{http://arxiv.org/abs/1611.08407}{{\tt arXiv:1611.08407}}].

\bibitem{Mazac:2018ycv}
D.~Mazac and M.~F. Paulos, {\it {The Analytic Functional Bootstrap II: Natural
  Bases for the Crossing Equation}},
  \href{http://arxiv.org/abs/1811.10646}{{\tt arXiv:1811.10646}}.

\bibitem{Mazac:2016qev}
D.~Mazac, {\it {Analytic bounds and emergence of AdS$_{2}$ physics from the
  conformal bootstrap}},  {\em JHEP} {\bf 04} (2017) 146,
  [\href{http://arxiv.org/abs/1611.10060}{{\tt arXiv:1611.10060}}].

\bibitem{Mazac:2018mdx}
D.~Mazac and M.~F. Paulos, {\it {The Analytic Functional Bootstrap I: 1D CFTs
  and 2D S-Matrices}},  \href{http://arxiv.org/abs/1803.10233}{{\tt
  arXiv:1803.10233}}.

\bibitem{ElShowk:2012hu}
S.~El-Showk and M.~F. Paulos, {\it {Bootstrapping Conformal Field Theories with
  the Extremal Functional Method}},  {\em Phys. Rev. Lett.} {\bf 111} (2013),
  no.~24 241601, [\href{http://arxiv.org/abs/1211.2810}{{\tt
  arXiv:1211.2810}}].

\bibitem{Hogervorst:2013sma}
M.~Hogervorst and S.~Rychkov, {\it {Radial Coordinates for Conformal Blocks}},
  {\em Phys. Rev.} {\bf D87} (2013) 106004,
  [\href{http://arxiv.org/abs/1303.1111}{{\tt arXiv:1303.1111}}].

\bibitem{Rychkov:2017tpc}
J.~Qiao and S.~Rychkov, {\it {Cut-touching linear functionals in the conformal
  bootstrap}},  {\em JHEP} {\bf 06} (2017) 076,
  [\href{http://arxiv.org/abs/1705.01357}{{\tt arXiv:1705.01357}}].

\bibitem{Pappadopulo:2012jk}
D.~Pappadopulo, S.~Rychkov, J.~Espin, and R.~Rattazzi, {\it {OPE Convergence in
  Conformal Field Theory}},  {\em Phys. Rev.} {\bf D86} (2012) 105043,
  [\href{http://arxiv.org/abs/1208.6449}{{\tt arXiv:1208.6449}}].

\bibitem{Qiao:2017xif}
J.~Qiao and S.~Rychkov, {\it {A tauberian theorem for the conformal
  bootstrap}},  {\em JHEP} {\bf 12} (2017) 119,
  [\href{http://arxiv.org/abs/1709.00008}{{\tt arXiv:1709.00008}}].

\bibitem{Hogervorst2017a}
M.~Hogervorst and B.~C. van Rees, {\it {Crossing symmetry in alpha space}},
  {\em JHEP} {\bf 11} (2017) 193, [\href{http://arxiv.org/abs/1702.08471}{{\tt
  arXiv:1702.08471}}].

\bibitem{Faller:2017hyt}
J.~Faller, S.~Sarkar, and M.~Verma, {\it {Mellin Amplitudes for Fermionic
  Conformal Correlators}},  {\em JHEP} {\bf 03} (2018) 106,
  [\href{http://arxiv.org/abs/1711.07929}{{\tt arXiv:1711.07929}}].

\bibitem{Zhou:2018sfz}
X.~Zhou, {\it {Recursion Relations in Witten Diagrams and Conformal Partial
  Waves}},  \href{http://arxiv.org/abs/1812.01006}{{\tt arXiv:1812.01006}}.

\bibitem{Liu:2018jhs}
J.~Liu, E.~Perlmutter, V.~Rosenhaus, and D.~Simmons-Duffin, {\it
  {$d$-dimensional SYK, AdS Loops, and $6j$ Symbols}},
  \href{http://arxiv.org/abs/1808.00612}{{\tt arXiv:1808.00612}}.

\bibitem{Sleight:2018ryu}
C.~Sleight and M.~Taronna, {\it {Anomalous Dimensions from Crossing Kernels}},
  {\em JHEP} {\bf 11} (2018) 089, [\href{http://arxiv.org/abs/1807.05941}{{\tt
  arXiv:1807.05941}}].

\bibitem{Gopakumar:2018xqi}
R.~Gopakumar and A.~Sinha, {\it {On the Polyakov-Mellin bootstrap}},
  \href{http://arxiv.org/abs/1809.10975}{{\tt arXiv:1809.10975}}.

\bibitem{Dey:2017fab}
P.~Dey, K.~Ghosh, and A.~Sinha, {\it {Simplifying large spin bootstrap in
  Mellin space}},  {\em JHEP} {\bf 01} (2018) 152,
  [\href{http://arxiv.org/abs/1709.06110}{{\tt arXiv:1709.06110}}].

\bibitem{Dey:2016mcs}
P.~Dey, A.~Kaviraj, and A.~Sinha, {\it {Mellin space bootstrap for global
  symmetry}},  {\em JHEP} {\bf 07} (2017) 019,
  [\href{http://arxiv.org/abs/1612.05032}{{\tt arXiv:1612.05032}}].

\bibitem{2016arXiv160304246V}
M.~{Viazovska}, {\it {The sphere packing problem in dimension 8}},  {\em ArXiv
  e-prints} (Mar., 2016) [\href{http://arxiv.org/abs/1603.04246}{{\tt
  arXiv:1603.04246}}].

\bibitem{2016arXiv160306518C}
H.~{Cohn}, A.~{Kumar}, S.~D. {Miller}, D.~{Radchenko}, and M.~{Viazovska}, {\it
  {The sphere packing problem in dimension 24}},  {\em ArXiv e-prints} (Mar.,
  2016) [\href{http://arxiv.org/abs/1603.06518}{{\tt arXiv:1603.06518}}].

\bibitem{ModularSphere}
D.~Mazac and L.~Rastelli, {\it {Modular Bootstrap and Sphere-Packing}},  {\em
  (in preparation)}.

\end{thebibliography}\endgroup
\bibliographystyle{jhep}

\end{document}